\documentclass[journal]{IEEEtran}
\usepackage{subfigure}
\usepackage{cite}
\usepackage{citesort}
\usepackage{amsmath,epsfig,bm}
\usepackage{pifont}
\usepackage{graphicx}
\usepackage{amssymb}
\usepackage{amsmath}
\usepackage{cancel}
\usepackage[normalem]{ulem}
\usepackage[dvips]{color}
\usepackage{algorithm}
\DeclareMathOperator*{\argmax}{argmax}

\usepackage{pifont}
\usepackage{graphicx}
\usepackage{amssymb}
\usepackage{amsmath}
\usepackage{cancel}
\usepackage[normalem]{ulem}
\usepackage[dvips]{color}
\usepackage{algorithm}
\usepackage{algorithmic}

\newtheorem{theorem}{\noindent \textbf{Theorem}}

\newcommand{\Hnull}{\mathcal{H}_0}
\newcommand{\Halt}{\mathcal{H}_1}
\newcommand{\Hk}{\mathcal{H}_k}
\newcommand{\Honull}{\mathcal{D}_0}
\newcommand{\Hoalt}{\mathcal{D}_1}

\begin{document}
\title{Location Verification Systems for VANETs in Rician Fading Channels}
\author{Shihao Yan, Robert Malaney, Ido Nevat, and Gareth W. Peters

\thanks{S. Yan and R. Malaney are with the School of Electrical Engineering and Telecommunications, The University of New South Wales, Sydney, NSW 2052, Australia (email: shihao.yan@unsw.edu.au; r.malaney@unsw.edu.au).}
\thanks{I. Nevat is with Institute for Infocomm Research, A$^{\star}$STAR, Singapore (email: ido-nevat@i2r.a-star.edu.sg).}
\thanks{G. W. Peters is with the Department of Statistical Science, University College London, London, United Kingdom (email: gareth.peters@ucl.ac.uk).}
\thanks{This work was funded by The University of New South Wales and Australian Research Council Grant DP120102607.}
}

\vspace{-6cm}

\maketitle

\begin{abstract}
In this work we propose and examine Location Verification Systems (LVSs) for Vehicular Ad Hoc Networks (VANETs) in the realistic setting of Rician fading channels. In our LVSs, a single authorized Base Station (BS) equipped with multiple antennas aims to detect a malicious vehicle that is spoofing its claimed location. We first determine the optimal attack strategy of the malicious vehicle, which in turn allows us to analyze the optimal LVS performance as a function of the  Rician $K$-factor of the channel between the BS and a legitimate vehicle. Our analysis also allows us to formally prove that the LVS performance limit is independent of the properties of the channel between the BS and the malicious vehicle, provided the malicious vehicle's antenna number is above a specified value. We also investigate how tracking information on a vehicle quantitatively improves the detection performance of an LVS, showing how optimal performance is obtained under the assumption of the tracking length being randomly selected. The work presented here can be readily extended to multiple BS scenarios, and therefore forms the foundation for all optimal location authentication schemes within the context of Rician fading channels. Our study closes important gaps in the current understanding of LVS performance within the context of VANETs, and will be of practical value to certificate revocation schemes within IEEE 1609.2.
\end{abstract}

\begin{keywords}
Location verification, location spoofing detection, Rician fading, likelihood ratio test, tracking information.
\end{keywords}

\section{Introduction}

In current wireless networks location-based techniques and services are now ubiquitous. As a consequence of this, the verification of location information has attracted considerable research interest in recent years \cite{malaney2004location,vora2006secure,chen2010detecting,chiang2012secure,zekavat2011handbook,yang2013detection,yan2014signal,yan2014optimal,malandrino2014verification}. In many location-based applications the device (client)  obtains its location information directly (e.g., via GPS),  and in such a case the wider network can only achieve the client's location  through requests to the client. In such a context, the client can easily spoof or falsify its claimed location in order to  disrupt some network functionalities  (e.g., geographic routing protocols \cite{leinmuller2005influence}, location-based access control protocols \cite{capkun2010integrity}). The adverse effects of
location spoofing can be more severe in Vehicular Ad Hoc Networks (VANETs) due to the possibility of life-threatening  accidents. Less critically, a malicious vehicle could spoof its location in order to seriously disrupt other vehicles \cite{raya2007securing}, or to selfishly enhance its own functionality within the network \cite{yu2013detecting}. The integrity of claimed location  in VANETs is therefore important, and motivates the introduction of a Location Verification System (LVS) to that scenario. Within IEEE 1609.2, an LVS will form part of the decision logic in the revocation of malicious-vehicle certificates  (see \cite{tang} for a review of certificate revocation within IEEE 1609.2).


Recently, many location verification protocols for VANETs have been proposed  (e.g.,\cite{leinmuller2006position,malaney2007wireless,yan2008providing,yan2009providing,hao2011cooperative,abumansoor2012a,zhang2012cooperative,yu2013detecting,fogue2014securing,yan2014location}).
These studies have proven useful in probing the detection performance of an LVS given a range of potential VANETs attack scenarios and an array of VANETs configurations. However, several important gaps in our knowledge of LVS performances and reliabilities remain. Among these are, (i) the optimal performance of an LVS as a function of the wireless channel conditions, and (ii) the optimal performance of an LVS as a function of the tracking information on a vehicle. These two open issues are of particular relevance to the VANETs environment, and the resolution of them forms the core of the work presented here.

With regard to our first issue, we note that
in VANETs environments Rician channels are anticipated to dominate the channel characteristics \cite{meireles2010experi,gozalves2012IEEE}. This fact allows us to specify more precisely the first question we wish to answer:
\emph{How does the optimal detection performance of an LVS quantitatively depend on the proportion of the LOS (line-of-sight) in a wireless channel?}
The proportion of the LOS in a wireless channel impacts the characteristics of observations obtained over wireless channels, such as the shadowing variance of Received Signal Strength (RSS), the estimation error of Time of Arrival (TOA), and the statistics on Angle of Arrival (AOA) determinations. The follow-on impact of such effects on LVS performances is non-trivial. Our approach in addressing this question will be to first determine the optimal attack strategy of the malicious vehicle, and to use that in order to conduct a formal theoretical analysis on the LVS performance.
With regard to our second issue, we pose the specific question: \emph{How does the tracking information on a vehicle quantitatively improve the detection performance of an LVS?} This question is of practical significance since under some  channel conditions the detection performance of an LVS with a only single claimed location (no tracking) is unfavorable. We address this issue by first formally developing the
optimal decision rule of an LVS via a Likelihood Ratio Test
(LRT) based on the track of claimed locations and then analyze the detection performance of an LVS when such tracking information is available.

In order to explicitly answer the above two questions, the directions and some specific contributions of this paper are summarized as follows. We first determine the optimal attack strategy of a malicious vehicle. To this end, after deriving the optimal transmit power and the optimal beamformer for the malicious vehicle at an arbitrary location, we identify the optimal locations of the malicious vehicle (best locations to launch an attack). Our analysis indicates that these optimal locations are determined solely by a single direction (due to the ability of the malicious vehicle to vary his transmit power and beamformer). Our analysis also reveals that the detection performance of an LVS will not be a function of the number of antennas held by the malicious vehicle once this number is above a derived bound.
We next establish that the optimal attack direction is that set by the direction from the claimed location to the BS (Base Station), and show how the malicious vehicle can perfectly imitate the signals expected from a legitimate vehicle if the malicious vehicle can find a location in this optimal direction with non-zero LOS.
However, given a constraint imposed that the true location of the malicious vehicle should be some minimum distance from its claimed location, such an optimal attack direction may not be viable. Considering unlimited resources possessed by the malicious vehicle (e.g., unlimited number of antennas), the LVS can determine the actual (now sub-optimal) best attack location given the constraint. We present how all of these findings allow us to establish lower bounds (worst-case scenario) on the detection performance of the LVS. We next extend our analysis to a tracking version of the LVS where multiple observations are utilized, showing how an extension of our previous analysis can lead to a range of similar outcomes, but with improved detection performance. A key part of the tracking LVS which allows for these findings is that the number of observations used for the decision-making process is randomly selected. Additional constraints on the tracking LVS solutions, imposed by speed limitations of the malicious vehicle, are presented. Finally, we present extensions of our analysis that take into account non-linear antenna arrays, and discuss the detection performance of the LVS in the presence of colluding attacks.

The rest of this paper is organized as follows. Section \ref{System Model} details our system model. In Section~\ref{sec_LVS-O}, the optimal attack strategy of the malicious vehicle is determined, based on which the detection performance of the LVS is analyzed. Section~\ref{sec_LVS-S} formalizes the optimal decision rule of the LVS when tracking information of the claimed location is available. In Section~\ref{sec_numerical}, we present numerical results to verify our analysis and we also draw some important insights based on our analysis. In Section~\ref{sec_discussion}, we discuss potential extension directions of our analysis and the impact of colluding attacks on an LVS. Finally, Section \ref{sec_conclusion} draws concluding remarks.

\emph{Notation:} Scalar variables are denoted by italic symbols. Vectors and matrices are denoted by lower-case and upper-case boldface symbols, respectively. Given a complex number $z$, $|z|$ denotes the modulus of $z$ and $\text{Re}\{z\}$ denotes the real part of $z$. Given a complex vector $\mathbf{x}$,
$\|\mathbf{x}\|$ denotes the Euclidean norm, $\mathbf{x}^{\top}$ denotes the transpose of $\mathbf{x}$, $\mathbf{x}^{\dag}$ denotes the conjugate transpose of $\mathbf{x}$, and $\mathbf{x}[i]$ denotes the $i$-th element of $\mathbf{x}$. Given a square matrix $\mathbf{X}$, $\text{tr}(\mathbf{X})$ denotes the trace of $\mathbf{X}$ and $\text{det}(\mathbf{X})$ denotes the determinant of $\mathbf{X}$. The $L\times L$ identity matrix is referred to as $\mathbf{I}_{L}$ and $\lceil \cdot \rceil$ denotes the ceiling function.

\section{System Model}\label{System Model}

\subsection{System Assumptions}

\begin{figure}[!t]
    \begin{center}
   {\includegraphics[width=3in]{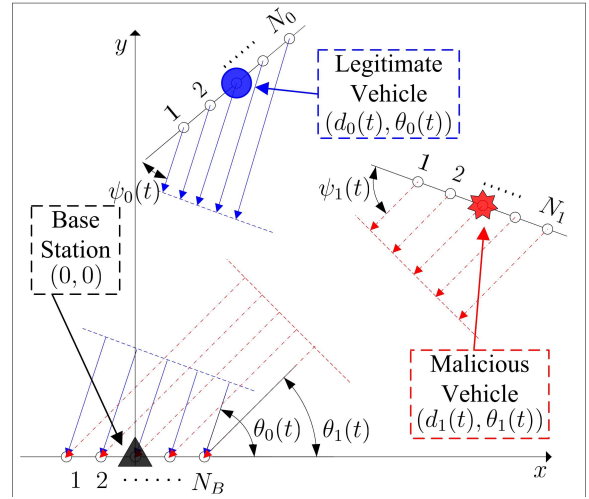}}
    \caption{Illustration of the orientations of the three ULAs and the geometry of the BS, the legitimate vehicle, and the malicious vehicle. We note that $N_1$, $d_1(t)$, $\theta_1(t)$, and $\psi_1(t)$ are not assumed to be known to the LVS.
}\label{fig:system}
    \end{center}
\end{figure}

Throughout this work we represent the inputs of an LVS as binary hypotheses, the null hypothesis $\Hnull$ and the alternative hypothesis $\Halt$. Under $\Hnull$ the vehicle is legitimate and provides to the LVS a claimed location equal to its true location. Under $\Halt$ the vehicle is malicious\footnote{Note, although we will often refer to the attacker as the \emph{malicious vehicle}, we should bear in mind that in reality the attacker may not be a vehicle (e.g., could be a generic device/user situated anywhere).} and  provides to the LVS a claimed location which is not its true location (a spoofed location).
We consider a VANETs application scenario, where the BS, the legitimate vehicle, and the malicious vehicle are all equipped with uniform linear arrays (ULAs). We discuss later the impact of non-linear antenna arrays. The number of antenna elements of the ULAs at the BS, the legitimate vehicle, and the malicious vehicle are $N_B$, $N_0$, and $N_1$, respectively.
Utilizing observations obtained over wireless channels, the BS is to verify whether the vehicle is indeed at its claimed location or not, thus inferring whether the vehicle is legitimate or malicious. In the first instance we will assume  the presence of only one malicious vehicle (we discuss colluding attacks later).

We adopt the polar coordinate system  ($d_k,\theta_k$) in this work ($k \in \{0,1\}$),  where $d_0$ ($d_1$) is the distance from the origin to the center of the legitimate (malicious) vehicle's ULA, and $\theta_0$ ($\theta_1$) represents the angle measured  counterclockwise from the $x$-axis to the line connecting the center of the legitimate (malicious) vehicle's ULA to the origin.  The location of the BS is selected as the origin, and the BS's ULA is aligned with the $x$-axis (antenna elements all on x-axis). A schematic of our assumed set-up is shown in Fig.~\ref{fig:system}. The claimed location of a vehicle (legitimate or malicious) at time slot $t$ ($t = 1, 2, \dots, T$) is denoted as $\mathbf{x}_c (t)=(d_c(t),\theta_c(t))$, which is supplied to the LVS  and to be verified (note, the LVS may be embedded in the BS). The true location of the vehicle under $\Hnull$ (the legitimate vehicle's true location) at $t$ is denoted as $\mathbf{x}_0(t) = [d_0(t), \theta_0(t)]^{\top}$. The true location of the vehicle under $\Halt$ (the malicious vehicle's true location) at $t$ is denoted as $\mathbf{x}_1(t) = [d_1(t), \theta_1(t)]^{\top}$.
Since the legitimate vehicle reports its true location to the LVS, we have $\mathbf{x}_c(t) = \mathbf{x}_0(t)$.
We adopt a practical threat model, in which the distance between the malicious vehicle's true location and its claimed location is larger than some specific value $r_l$ (i.e., $\|\mathbf{x}_1 (t) - \mathbf{x}_c (t)\| > r_l$).
We note that this assumption is reasonable since the malicious vehicle does not need to spoof its claimed location if $\|\mathbf{x}_1 (t) - \mathbf{x}_c (t)\|$ is very small. The value of $r_l$ can be predetermined based on some specific application scenario and in general it is larger than a vehicle's intrinsic position uncertainty. The angles $\psi_0(t)$ and $\psi_1(t)$ as shown in Fig.~\ref{fig:system} are under the control of the legitimate and malicious vehicles, respectively. We note that $\psi_0(t)$ ($\psi_1(t)$) represents the angle measured counterclockwise from the orientation of the ULA at the legitimate (malicious) vehicle to the line connecting the center of the legitimate (malicious) vehicle's ULA to the origin. Without other statements, we assume all information available to the LVS, BS, and legitimate vehicle is also known to the malicious vehicle. We  assume $N_1$, $\mathbf{x}_1 (t)$, and $\psi_1(t)$ are known only by the malicious vehicle. We will assume that $N_1$ is unbounded (of course in practice this number is constrained by the communication wavelength and the physical dimensions of the vehicle). If in practice the  malicious vehicle possesses less than a critical number of antenna elements (to be derived later), then the results presented here represent conservative lower bounds on the LVS performance.

Note in this work we will consider observations collected by only one BS. In general, this represents the most likely (default) scenario for many real-world VANETs.  As such, the analysis we provide here should be widely applicable. The analysis for the single BS also forms the basis from which other more complicated scenarios can be built upon.  For example, in instances where  additional BSs are within range of claimed positions, the work presented here can be readily adapted to account for that.\footnote {We note that other trusted vehicles within range of the claimed position could also be used as additional reference stations. Indeed, vehicles which are considered legitimate (e.g., by consistently passing all LVS decisions over a length of time) can be used to dynamically create/update the $K$ map for a particular BS, at least with regard to all locations on the road.} A conceptually simple method of doing this would be for each additional BS to be allocated a separate LVS which can then cooperate  with other LVSs (BSs) in order to make optimally-joint decisions.

\subsection{Channel Model}

We assume the channel from a vehicle (legitimate or malicious) to the BS is subject to Rician fading. Then, the $N_B \times N_k$ channel matrix  at $t$ under $\Hk$ is given by
\begin{align}\label{h_definition}
\mathbf{H}_k(t) = \sqrt{\frac{K_k(t)}{1+K_k(t)}} \overline{\mathbf{H}}_k(t) + \sqrt{\frac{1}{1+K_k(t)}} \widetilde{\mathbf{H}}_k(t),
\end{align}
where $K_k(t)$ is the Rician $K$-factor of the channel under $\Hk$ (we assume $K_k(t)$ is a function the vehicle's true location), $\overline{\mathbf{H}}_k(t)$ is the LOS component of $\mathbf{H}_k(t)$, and $\widetilde{\mathbf{H}}_k(t)$ is the scattered component of $\mathbf{H}_k(t)$. The entries of $\widetilde{\mathbf{H}}_k(t)$ are independent and identically distributed (i.i.d.) circularly-symmetric complex Gaussian random variables with zero mean and unit variance. We assume that $\widetilde{\mathbf{H}}_k(t)$ is i.i.d. in different time slots. Denoting $\rho_B$ as the space between two adjacent antenna elements of the ULA at the BS, $\overline{\mathbf{H}}_k(t)$ can be written as
$\overline{\mathbf{H}}_k(t) = \mathbf{r}_{k}(t)  \mathbf{t}_k(t)$ \cite{taricco2011on},
where $\mathbf{r}_k(t)$ and $\mathbf{t}_k(t)$ are given by
\begin{align}
\mathbf{r}_{k}(t) &= \left[1,\cdots,\exp(j (N_B -1)\tau_B \cos \theta_k(t))\right]^{\top},\label{r_L_definition} \\
\mathbf{t}_k(t)  &= \left[1,\cdots,\exp(-j (N_k -1)\tau_k \cos \psi_k(t))\right]. \label{t_L_definition}
\end{align}
In \eqref{r_L_definition} and \eqref{t_L_definition}, we have $\tau_B = 2 \pi f_c \rho_B /c$ and $\tau_k = 2 \pi f_c \rho_k /c$, where $f_c$ is the carrier frequency, $c$ is the speed of propagation of the plane wave, $\rho_0$ is the space between two antenna elements of the ULA at the legitimate vehicle, and $\rho_1$ is the space between two antenna elements of the ULA at the malicious vehicle. We note that we assume the LVS knows $K_0(t)$ (e.g., through a predetermined measurement campaign in the vicinity of the BS). We assume $K_1(t)$ is known by the malicious vehicle but not known by the LVS. Note that we will assume that the time dependence for all our variables arises solely from the fact that the vehicle is in general moving (i.e., the variables are functions of location). The exception to this is $\widetilde{\mathbf{H}}_k(t)$, for which the time dependence is also due to the movement of scatterers. Our channel model covers the entire range of conditions from a pure Rayleigh channel ($K=0$) to a pure LOS channel ($K=\infty$).

\subsection{Observation Model}

The composite observation model is given by
\begin{align}\label{composite_model}
\Hk: \mathbf{y}(t) \!=\! \sqrt{p_k(t) \mathrm{g}(d_k(t))}\mathbf{H}_k(t)\mathbf{b}_k(t) s \!+\! \mathbf{n}_k(t),
\end{align}
where $p_k(t)$ is the transmit power of the vehicle under $\Hk$,\footnote{We will be conservative and  assume the attacker has unlimited power resources. If a power constraint (on attacker) is introduced some of the attacks we describe later may not be possible, and in these circumstances the LVS performances shown can be considered lower bounds (worst-case scenarios).} $\mathrm{g}(d_k(t))$ is the path loss gain under $\Hk$ given by $\mathrm{g}(d_k(t)) = \left(c/4\pi f_c d_r\right)^2({d_r}/{d_k(t)})^{\xi}$, $d_r$ is a reference distance, $\xi$ is the path loss exponent, $\mathbf{b}_k(t)$ is the beamformer adopted by the vehicle under $\Hk$ that satisfies $\|\mathbf{b}_k(t)\| = 1$, $s$ is the publicly known pilot symbol satisfying $\|s\| = 1$, and $\mathbf{n}_k(t)$ is the additive white Gaussian noise vector at $t$ under $\Hk$, of which the entries are i.i.d circularly-symmetric complex Gaussian random variables with zero mean and variance $\sigma_k^2$. We note that for simplicity we assume that $\xi$ is independent of a vehicle's location and is the same for all power components (i.e., $\overline{\mathbf{H}}_k(t)$ and $\widetilde{\mathbf{H}}_k(t)$) since we have assumed that $K_k(t)$ is a function of a vehicle's location \cite{prasad1993effects}. As we show later, our analysis still holds even if $\xi$ is a function of a vehicle's location and is different for $\overline{\mathbf{H}}_k(t)$ and $\widetilde{\mathbf{H}}_k(t)$. We assume that the legitimate vehicle adopts constant transmit power, i.e., $p_0(t) = p_0$. However, we note that $p_1(t)$ varies. This is due to the fact that the malicious vehicle can adjust its transmit power based on each pair of $\mathbf{x}_0(t)$ and $\mathbf{x}_1(t)$.
We also assume that $\mathbf{n}_k(t)$ is i.i.d in different time slots.
We note that $\mathbf{b}_0(t)$ and $p_0$ are under the control of the legitimate vehicle. We assume that the legitimate vehicle cooperates with the BS to facilitate the location verification. To this end, the legitimate vehicle sets $\mathbf{b}_0(t) = \mathbf{t}_0^{\dag}(t)/\|\mathbf{t}_0(t) \|$ so as to maximize $|\mathbf{t}_0(t) \mathbf{b}_0(t)|$. In addition, the legitimate vehicle sets its transmit power to the required value by the BS (we assume $p_0$ is publicly known).  Again, we assume neither $p_1(t)$ nor $\mathbf{b}_1(t)$ is known to the LVS.
 According to \eqref{h_definition} and \eqref{composite_model}, the likelihood function of $\mathbf{y}(t)$ conditioned on a known $s$ under $\Hk$ is
\begin{align}\label{pdf_y0}
f&(\mathbf{y}(t)|\Hk) \!=\! \frac{1}{\pi^{N_B}\det(\mathbf{R}_k(t))} \times \notag \\ &~~~~~~~~\exp\left[\!-\!(\mathbf{y}(t)\!\!-\!\!\mathbf{m}_k(t))^{\dag}\mathbf{R}_k^{\!-\!1}(t)(\mathbf{y}(t)\!\!-\!\!\mathbf{m}_k(t))\right],
\end{align}
where $\mathbf{m}_k(t)$ and $\mathbf{R}_k(t)$ are the mean vector and covariance matrix of $\mathbf{y}(t)$ under $\Hk$, respectively, which are given by
\begin{align}
\mathbf{m}_k(t) &= \sqrt{\frac{p_k(t) \mathrm{g}(d_k(t)) K_k(t) }{1+K_k(t)}}\overline{\mathbf{H}}_k(t)\mathbf{b}_k(t),\label{mk_definition}\\
\mathbf{R}_k(t) &= \left(\frac{p_k(t) \mathrm{g}(d_k(t))}{1 + K_k(t)} + \sigma_k^2\right) \mathbf{I}_{N_B}.\label{R0_definition}
\end{align}

We note that under $\Hnull$ we have $\overline{\mathbf{H}}_0(t)\mathbf{b}_0(t) = \sqrt{N_0} \mathbf{r}_0(t)$ due to $\mathbf{b}_0(t) = \mathbf{t}_0^{\dag}(t)/\|\mathbf{t}_0(t) \|$. We also note that $f(\mathbf{y}(t)|\Halt)$ is dependent on $p_1(t)$, $\mathbf{b}_1(t)$, and $\mathbf{x}_1(t)$. Thus, we also denote $f(\mathbf{y}(t)|\Halt)$ as
$f\left(\mathbf{y}|p_1(t), \mathbf{b}_1(t), \mathbf{x}_1(t), \Halt\right)$. These parameters (i.e., $p_1(t)$, $\mathbf{b}_1(t)$, and $\mathbf{x}_1(t)$) are all under the control of the malicious vehicle and are unknown to the LVS. In the next section, we will discuss how the malicious vehicle optimally sets these parameters in order to minimize the probability of being detected by the LVS.

\section{Location Verification System \\Without Tracking}\label{sec_LVS-O}

In this section we examine the performance of the LVS by considering only one claimed location and one observation snapshot at the BS antennas (i.e., BS measurements made in one time slot, and $T = 1$). As such, we drop explicit reference to $(t)$ for all variables in this section. We first present the decision rule and performance metrics adopted in this LVS. We then discuss the optimal attack strategy of the malicious vehicle (i.e., how to optimally set $p_1$, $\mathbf{b}_1$, and $\mathbf{x}_1$) in order to minimize the probability to be detected. Finally, we analyze the detection performance of the LVS based on this optimal attack strategy.

\subsection{Decision Rule of the LVS}

We adopt the LRT as the decision rule of the LVS. This is due to the fact that the LRT achieves the highest detection rate (the probability to correctly detect a malicious vehicle) for any given false positive rate (the probability to incorrectly detect a legitimate vehicle as malicious) \cite{neyman1933problem}. The LRT decision rule is given by
\begin{equation}\label{arbitrary}
\Lambda\left(\mathbf{y}\right) \triangleq \frac{f\left(\mathbf{y}|p_1, \mathbf{b}_1, \mathbf{x}_1, \Halt\right)}{f\left(\mathbf{y}|\Hnull\right)} \begin{array}{c}
\overset{\Hoalt}{\geq} \\
\underset{\Honull}{<}
\end{array}%
\lambda,
\end{equation}
where $\Lambda\left(\mathbf{y}\right)$ is the likelihood ratio of $\mathbf{y}$, $\lambda$ is the threshold for $\Lambda\left(\mathbf{y}\right)$, and $\Honull$ and $\Hoalt$ are the binary decisions that infer whether the vehicle is legitimate or malicious, respectively. Given the decision rule in \eqref{arbitrary}, the false positive and detection rates of the LVS are functions of $\lambda$. Note, the false positive rate is given by $\alpha(\lambda) = \Pr\left(\Lambda\left(\mathbf{y}\right) > \lambda|\Hnull\right)$, and detection rate is given by  $\beta(\lambda) =\Pr\left(\Lambda\left(\mathbf{y}\right) > \lambda|\Halt\right)$.
The specific value of $\lambda$ can be set through predetermining a false positive rate, minimizing the Bayesian average cost, or maximizing the mutual information between the system input and output \cite{yan2014optimal}. In order to quantitatively examine the impact of some system parameters on the detection performance of the LVS, we have to adopt a unique metric to evaluate the LVS. When it is necessary, we adopt a special Bayesian average cost as the unique performance metric, which is the total error. The total error is obtained by setting the costs of correct and incorrect decisions as zeros and ones, respectively \cite{barkat2005signal}. The total error can be expressed as
\begin{align}\label{total_error}
\epsilon(\lambda) = P_0 \alpha(\lambda) + (1-P_0) (1 - \beta(\lambda)),
\end{align}
where $P_0$ and $1-P_0$ are the \emph{a priori} probabilities that the vehicle is legitimate and malicious, respectively. Based on the Bayesian framework, the optimal value of $\lambda$ that minimizes $\epsilon(\lambda)$ is given by $\lambda^{\ast} = P_0/(1-P_0)$ \cite{barkat2005signal}. Substituting $\lambda^{\ast}$ into \eqref{total_error}, we can obtain the minimum value of $\epsilon(\lambda)$, referred to as the \emph{minimum total error} and denoted by $\epsilon^{\ast}$.


\subsection{Optimal Attack Strategy Against the LVS}

Knowing \eqref{arbitrary}, the malicious vehicle is to minimize the difference between $f\left(\mathbf{y}|p_1, \mathbf{b}_1, \mathbf{x}_1, \Halt\right)$ and $f\left(\mathbf{y}|\Hnull\right)$ in order to minimize the detection rate.
It can be shown that minimization of
the Kullback-Leibler (KL) divergence leads to the minimum detection rate \cite{eguchi2006interpreting}. This is due to that the KL divergence is also the expected log likelihood ratio when the alternative hypothesis $\Halt$ is true. The KL divergence from $f\left(\mathbf{y}|p_1, \mathbf{b}_1, \mathbf{x}_1, \Halt\right)$ to $f\left(\mathbf{y}|\Hnull\right)$ is defined as \cite{kullback1951on}
\begin{align}
&D_{KL}\left(f\left(\mathbf{y}|p_1, \mathbf{b}_1, \mathbf{x}_1, \Halt\right)||f\left(\mathbf{y}|\Hnull\right)\right) \notag \\
&~~~~~~~~~~~~~~~~= \int \left[\ln  \Lambda (\mathbf{y}) \right] f\left(\mathbf{y}|p_1, \mathbf{b}_1, \mathbf{x}_1, \Halt\right) d \mathbf{y}. \label{KL_definition}
\end{align}
Given this, the optimization problem for the malicious vehicle can be written as
\begin{equation}\label{optimize_m}
\begin{split}
&\left(p_1, \mathbf{b}_1, \mathbf{x}_1\right)^{\ast} = \\
&\quad \argmax_{p_1 \geq 0, \|\mathbf{b}_1\| = 1, \atop \|\mathbf{x}_c - \mathbf{x}_1\| \geq r_l} D_{KL}\left(f\left(\mathbf{y}|p_1, \mathbf{b}_1, \mathbf{x}_1, \Halt\right)||f\left(\mathbf{y}|\Hnull\right)\right).
\end{split}
\end{equation}
We present the solutions to \eqref{optimize_m} in two steps. We first derive the optimal values of $p_1$ and $\mathbf{b}_1$ for any given $\mathbf{x}_1$ in Theorem~\ref{theorem1}. Then, we search for the optimal value of $\mathbf{x}_1$ numerically, with the aid of Theorem~\ref{theorem2}.

\begin{theorem}\label{theorem1}
The optimal values of $p_1$ and $\mathbf{b}_1$ that minimize the detection rate for any given $\mathbf{x}_1$ are derived as
\begin{align}
p_1^{\ast}(\mathbf{x}_1) &= \frac{K_1 + 1}{g(d_1)}\left(\frac{p_0 g(d_0)}{1+K_0} + \sigma_0^2 - \sigma_1^2\right),\label{opt_P1}\\
\mathbf{b}_1^{\ast}(\mathbf{x}_1) &= \mathbf{U}_{\ast} \mathbf{p}^{\ast}, \label{opt_b1}
\end{align}
where $\mathbf{U}_{\ast}$ is the left singular and orthogonal matrix of the Singular Value Decomposition (SVD) for $\mathbf{G}_{\ast}^{\dag}\mathbf{G}_{\ast}$, $\mathbf{G}_{\ast} = \sqrt{{p_1^{\ast}(\mathbf{x}_1) \mathrm{g}(d_1) K_1}/(1+K_1)} ~\overline{\mathbf{H}}_1$, $\mathbf{p}^{\ast}[1] = \mathbf{U}_{\ast}^{\dag}\mathbf{G}_{\ast}^{\dag}\mathbf{m}_0[1]/\eta_{\ast}$,  $\eta_{\ast}$ is the unique eigenvalue of $\mathbf{G}_{\ast}^{\dag}\mathbf{G}_{\ast}$, and $\mathbf{p}^{\ast}[i]$ for $i = 2, 3, \cdots, N_1$ can be any value which enables $\|\mathbf{p}^{\ast}\|=1$,
\end{theorem}
\begin{IEEEproof}
Substituting \eqref{pdf_y0} into \eqref{KL_definition}, we have
\begin{equation}\label{KL_result}
\begin{split}
D_{KL}&\left(f\left(\mathbf{y}|p_1, \mathbf{b}_1, \mathbf{x}_1, \Halt\right)||f\left(\mathbf{y}|\Hnull\right)\right) \\
&~~~~~~~~= \underbrace{\text{tr}(\mathbf{R}_0^{-1}\mathbf{R}_1) \!-\!N_B\!-\! \ln \left(\frac{\det \mathbf{R}_1}{\det \mathbf{R}_0}\right)}_{h_1(p_1)} \\
&~~~~~~~~~~~~+ \underbrace{(\mathbf{m}_0 \!-\! \mathbf{m}_1)^{\dag}\mathbf{R}_0^{-1}(\mathbf{m}_0 \!-\! \mathbf{m}_1)}_{h_2(p_1,\mathbf{b}_1)}.
\end{split}
\end{equation}
Based on \eqref{KL_result}, we know that only the term $h_2\left(p_1, \mathbf{b}_1\right)$ is a function of $\mathbf{b}_1$. As such, we first derive the optimal $\mathbf{b}_1$ that minimizes $h_2\left(p_1, \mathbf{b}_1\right)$ for a given $p_1$. Given the format of $\mathbf{R}_0$ presented in \eqref{R0_definition}, we can see that $h_2\left(p_1, \mathbf{b}_1\right)$ is minimized when $\|\mathbf{m}_0 - \mathbf{m}_1\|^2$ is minimized. Defining $\mathbf{G} = \sqrt{{p_1 \mathrm{g}(d_1) K_1}/(1+K_1)} ~\overline{\mathbf{H}}_1$, we have
\begin{align}\label{h3}
h_3\left(\mathbf{b}_1\right) &\triangleq  \|\mathbf{m}_0 - \mathbf{m}_1\|^2 \notag \\
&= \mathbf{b}_1^{\dag}\mathbf{G}^{\dag}\mathbf{G}\mathbf{b}_1 \!-\! \mathbf{m}_0^{\dag}\mathbf{G} \mathbf{b}_1 \!-\! \mathbf{b}_1^{\dag}\mathbf{G}^{\dag} \mathbf{m}_0 \!+\! \mathbf{m}_0^{\dag}\mathbf{m}_0.
\end{align}
Performing the SVD for the symmetric positive semidefinite matrix $\mathbf{Q} \triangleq \mathbf{G}^{\dag}\mathbf{G}$, we have
\begin{align}\label{svd_q}
\mathbf{U}\mathbf{V}\mathbf{U}^{\dag} = \mathbf{Q}.
\end{align}
We note that $\mathbf{Q}$ is a rank-1 matrix and we denote the unique eigenvalue of $\mathbf{Q}$ as $\eta$. Then, we have
\begin{align}\label{eta_d}
\eta = \|\mathbf{Q}\| = \frac{p_1 g(d_1) K_1 N_B N_1}{1+K_1}.
\end{align}
Denoting $\mathbf{b}_1 = \mathbf{U} \mathbf{p}$ (i.e., $\mathbf{p} = \mathbf{U}^{\dag} \mathbf{b}_1$), following \eqref{h3} and \eqref{svd_q} we have
\begin{align}\label{h3_p}
h_3\left(\mathbf{b}_1\right) \!=\! \mathbf{p}^{\dag}\mathbf{V}\mathbf{p} \!-\! \mathbf{m}_0^{\dag}\mathbf{G} \mathbf{U} \mathbf{p} \!-\! \mathbf{p}^{\dag} \mathbf{U}^{\dag}\mathbf{G}^{\dag} \mathbf{m}_0 + \mathbf{m}_0^{\dag}\mathbf{m}_0.
\end{align}
We note that $\mathbf{U}^{\dag}\mathbf{G}^{\dag} \mathbf{m}_0$ is a complex $N_1 \times 1$ vector and we denote the $i$-th complex element of $\mathbf{U}^{\dag}\mathbf{G}^{\dag} \mathbf{m}_0$ as $c_{Ri} + j c_{Ii}$. Since $\mathbf{Q}$ is a rank-1 matrix, we have $\mathbf{U}^{\dag}\mathbf{G}^{\dag} \mathbf{m}_0[i] = 0$ for $i = 2, 3, \cdots, N_1$. Denoting the $i$-th complex element of $\mathbf{p}$ as $p_{Ri} + j p_{Ii}$, following \eqref{h3_p} we have
\begin{align}\label{h_3_pf}
h_3\left(\mathbf{b}_1\right) \!=\! \eta (p_{R1}^2 \!+\! p_{I1}^2) \!-\! 2 \left(c_{R1}p_{R1} \!+\! c_{I1}p_{I1}\right) + \mathbf{m}_0^{\dag}\mathbf{m}_0.
\end{align}
Using \eqref{h_3_pf}, we have $p_{R1} = c_{R1}\eta$ and $p_{I1} = c_{I1}\eta$ in order to minimize $h_3\left(\mathbf{b}_1\right)$ without any constraints, which results in
\begin{align}\label{po_opt}
\mathbf{p}^o[1] = \mathbf{U}^{\dag}\mathbf{G}^{\dag}\mathbf{m}_0[1]/\eta,
\end{align}
where $\mathbf{p}^o$ denotes the optimal $\mathbf{p}$ that minimizes $h_3(\mathbf{p})$ for a given $p_1$.
We note that there is a constraint for the minimization of $h_3\left(\mathbf{b}_1\right)$, which is  $\|\mathbf{b}_1\| = 1$ (i.e., $\|\mathbf{p}\| = 1$ since $\mathbf{U}$ is a unitary matrix). As such, we have to guarantee $c_{R1}^2 \!+\! c_{I1}^2 \leq \eta$, which means that we have to guarantee $\|\mathbf{U}^{\dag}\mathbf{G}^{\dag}\mathbf{m}_0\|/\eta \leq 1$. Based on the definitions of $\mathbf{G}$ and $\mathbf{m}_0$, and noting $\mathbf{t}_1\mathbf{t}_1^{\dag} = N_1$ we have
\begin{align}\label{N1condition}
\|\mathbf{U}^{\dag}\mathbf{G}^{\dag}\mathbf{m}_0\|^2 &= \|\mathbf{G}^{\dag}\mathbf{m}_0\|^2\notag \\
&= {\frac{p_0 \mathrm{g}(d_0) K_0 N_0 }{1+K_0}} {\frac{p_1 \mathrm{g}(d_1) K_1 }{1+K_1}} \mathbf{r}_0^{\dag}\mathbf{r}_1 \mathbf{t}_1\mathbf{t}_1^{\dag}
\mathbf{r}_1^{\dag}\mathbf{r}_0 \notag \\
&= {\frac{p_0 \mathrm{g}(d_0) K_0 N_0 }{1+K_0}} {\frac{p_1 \mathrm{g}(d_1) K_1 N_1}{1+K_1}}
|\mathbf{r}_1^{\dag}\mathbf{r}_0|^2.
\end{align}
We also note that the maximum value of $|\mathbf{r}_1^{\dag}\mathbf{r}_0|^2$ is $N_B^2$, which is achieved when $\mathbf{r}_1 = \mathbf{r}_0$. Then, as per \eqref{eta_d} we have
\begin{align}\label{N1condition}
\frac{\|\mathbf{U}^{\dag}\mathbf{G}^{\dag}\mathbf{m}_0\|}{\eta} \leq \underbrace{\sqrt{\frac{p_0 \mathrm{g}(d_0) K_0 }{1+K_0}} \sqrt{\frac{1+K_1}{p_1 \mathrm{g}(d_1) K_1 }} \sqrt{\frac{N_0}{N_1}}}_{\mathcal{L}(N_1)}.
\end{align}
In order to guarantee $\mathcal{L}(N_1) \leq 1$, the malicious vehicle has to guarantee $N_1 \geq N_1^{\ast}$, where $N_1^{\ast}$ is obtained by setting $\mathcal{L}(N_1) = 1$ and is given by
\begin{align}
N_1^{\ast} = \left\lceil \max \left\{2,\frac{p_0 g(d_0) K_0 N_0}{K_1[p_0 g(d_0) \!+\! (1\!+\!K_0)(\sigma_0^2 \!-\! \sigma_1^2)]}\right\}\right\rceil \label{opt_N1}.
\end{align}
The reason for $N_1^{\ast} \geq 2$ is that the minimum dimension of $\mathbf{p}$ must be $2$ if $\mathbf{r}_1$ is to remain a function of $\theta_1$. We assume the malicious vehicle can  guarantee $N_1 \geq N_1^{\ast}$, and therefore guarantee  $\|\mathbf{U}^{\dag}\mathbf{G}^{\dag}\mathbf{m}_0\|/\eta \leq 1$.
As such, the optimal solution $\mathbf{p}^o[1] = \mathbf{U}^{\dag}\mathbf{G}^{\dag}\mathbf{m}_0[1]/\eta$ can always be achieved. This optimal solution indicates that $\mathbf{p}^o[i]$, for $i \geq 2$, can take any values in order to  realize $\|\mathbf{p}^o\| = 1$.

We next derive the optimal value of $p_1$. Substituting $\mathbf{p}^o[1] = \mathbf{U}^{\dag}\mathbf{G}^{\dag}\mathbf{m}_0[1]/\eta$ into \eqref{h3_p}, we have
\begin{align}\label{h3_po}
h_3(\mathbf{b}_1^o) &\!=\! \mathbf{m}_0^{\dag}\mathbf{m}_0 \!-\! \frac{\|\mathbf{U}^{\dag}\mathbf{G}^{\dag}\mathbf{m}_0\|^2}{\eta}\notag\\
&= \frac{p_0 g(d_0)K_0 N_0}{1+K_0} \left( N_B - \frac{| \mathbf{r}_1^{\dag}\mathbf{r}_0|^2}{N_B}\right),
\end{align}
where $\mathbf{b}_1^o = \mathbf{U} \mathbf{p}^o$.
We note that $|\mathbf{r}_1^{\dag}\mathbf{r}_0|^2$ is a function of only $N_B$, $\theta_0$, and $\theta_1$. Thus, $h_3(\mathbf{b}_1^o)$ is not a function of $p_1$ anymore. Based on \eqref{KL_result}, we know that $h_1(p_1)$ is a function of only $p_1$. This indicates that the optimal $p_1$ is the one that minimizes $h_1(p_1)$. After some algebra, we can show that that $h_1(p_1)$ is minimized when $\mathbf{R}_0 = \mathbf{R}_1$, which results in the desirable result in \eqref{opt_P1}. We note that to achieve \eqref{opt_P1} we require $\sigma_1^2 < p_0 g(d_0)/(1+K_0) + \sigma_0^2$. This is reasonable as the channel noise variance will be lower than the useful signal power.
Finally, substituting $p_1^{\ast}(\mathbf{x}_1)$ into \eqref{po_opt} we obtain the desirable result in \eqref{opt_b1}.
\end{IEEEproof}

We note that if the condition $N_1 \geq N_1^{\ast}$ cannot be guaranteed, the minimum KL divergence for any given $\mathbf{x}_1$ will be larger than that for $N_1 \geq N_1^{\ast}$. To prove this statement, we have to prove the following equation
\begin{equation}\label{N1_less1}
\begin{split}
&D_{KL}\left(f\left(\mathbf{y}|p_1^{\prime}(\mathbf{x}_1), \mathbf{b}_1^{\prime}(\mathbf{x}_1), \mathbf{x}_1, \Halt\right)||f\left(\mathbf{y}|\Hnull\right)\right)  \\
& \quad \geq D_{KL}\left(f\left(\mathbf{y}|p_1^{\ast}(\mathbf{x}_1), \mathbf{b}_1^{\ast}(\mathbf{x}_1), \mathbf{x}_1, \Halt\right)||f\left(\mathbf{y}|\Hnull\right)\right),
\end{split}
\end{equation}
where $p_1^{\prime}(\mathbf{x}_1)$ and $\mathbf{b}_1^{\prime}(\mathbf{x}_1)$ denote the optimal values of $p_1$ and $\mathbf{b}_1$ under the condition $N_1 < N_1^{\ast}$ for any given $\mathbf{x}_1$.
Following \eqref{h_3_pf}, we have $h_3(\mathbf{b}_1^{\prime}(\mathbf{x}_1)) \geq h_3(\mathbf{b}_1^{\ast}(\mathbf{x}_1))$. This is due to the fact that $\mathbf{b}_1^{\prime}(\mathbf{x}_1)$ minimizes $h_3(\mathbf{b}_1)$ under the constraint $p_{R1}^2 \!+\! p_{I1}^2 \leq 1$, but $\mathbf{b}_1^{\ast}(\mathbf{x}_1)$ minimizes $h_3(\mathbf{b}_1)$ without any constraints. Noting $h_1(p_1^{\ast}(\mathbf{x}_1)) = 0$, we have $h_1(p_1^{\prime}(\mathbf{x}_1)) \geq h_1(p_1^{\ast}(\mathbf{x}_1))$. This is due to $h_1(p_1) \geq 0$ for any values of $p_1$ since the KL divergence is not negative. Then, we have
\begin{align}\label{N1_less2}
h_1(p_1^{\prime}(\mathbf{x}_1)) \!+\! h_3(\mathbf{b}_1^{\prime}(\mathbf{x}_1)) \!\geq\! h_1(p_1^{\ast}(\mathbf{x}_1)) \!+\! h_3(\mathbf{b}_1^{\ast}(\mathbf{x}_1)).
\end{align}
Since $\mathbf{R}_0$ is independent of $N_1$, following \eqref{KL_result} we can see  \eqref{N1_less2} proves \eqref{N1_less1}.

\begin{theorem}\label{theorem2}
The optimal value of $\theta_1$ that minimizes the detection rate can be obtained through
\begin{align}\label{opt_x1}
\theta_1^{\ast} = \argmax_{\|\mathbf{x}_c - \mathbf{x}_1\| \geq r_l} |\mathbf{r}_1^{\dag}\mathbf{r}_0|^2.
\end{align}
\end{theorem}

\begin{IEEEproof}
Substituting \eqref{opt_P1} and \eqref{opt_b1} into \eqref{KL_result}, we obtain the minimum value of $D_{KL}\left(f\left(\mathbf{y}|p_1, \mathbf{b}_1, \mathbf{x}_1, \Halt\right)||f\left(\mathbf{y}|\Hnull\right)\right)$ for any given $\mathbf{x}_1$ as
\begin{align}\label{KL_minimum_xt}
&D_{KL}\left(f\left(\mathbf{y}|p_1^{\ast}(\mathbf{x}_1), \mathbf{b}_1^{\ast}(\mathbf{x}_1), \mathbf{x}_1, \Halt\right)||f\left(\mathbf{y}|\Hnull\right)\right)\notag \\
&~~~~~~~~= \frac{p_0 g(d_0)K_0 N_0}{p_0 g(d_0) + \sigma_0^2 (1+K_0)} \left( N_B - \frac{|\mathbf{r}_1^{\dag}\mathbf{r}_0|^2}{N_B}\right).
\end{align}
The malicious vehicle will determine its optimal true location by finding the value of $\mathbf{x}_1$ that minimizes \eqref{KL_minimum_xt}. We note that in \eqref{KL_minimum_xt} only the term $|\mathbf{r}_1^{\dag}\mathbf{r}_0|^2$ is a function of $\theta_1$. As such,  the malicious vehicle needs only to maximize $|\mathbf{r}_1^{\dag}\mathbf{r}_0|^2$ in order find the optimal $\theta_1$. As such, we obtain  \eqref{opt_x1}.
\end{IEEEproof}

Based on Theorem~\ref{theorem1} and Theorem~\ref{theorem2} we obtain the following important insights.
(i) We note that once $N_1 = N_1^{\ast}$, further increases in $N_1$ offer no further benefit to the malicious vehicle. That is, the additional degrees of freedom offered by additional antennas beyond $N_1^{\ast}$  serve no purpose (in the beamformer solution the malicious vehicle can set power allocated to these additional antennas - if it has them - to zero).
(ii) We can see that the minimum KL divergence presented in \eqref{KL_minimum_xt} increases as $p_0$, $g(d_0)$, $K_0$, or $N_0$ increases.
(iii) We note that the minimum KL divergence presented in \eqref{KL_minimum_xt} is zero when $K_0 = 0$, and thus the malicious vehicle can always perfectly imitate the legitimate vehicle (again this issue that could be neutralized by using additional BSs).
(iv) We note that the minimum KL divergence provided in \eqref{KL_minimum_xt} is not a function of
$K_1$ or $\sigma_1^2$. However, we  highlight that as $K_1\rightarrow 0$, $N_1^{\ast} \rightarrow \infty$, meaning $K_1 = 0$ represents the worst case for the malicious vehicle.
(v) Based on Theorem~\ref{theorem2} we note that $\theta_1^{\ast}$ is a function of only $\mathbf{r}_0$ (i.e., only depends on $N_B$ and $\theta_0$). This indicates that the malicious vehicle can directly search for its true location as per Theorem~\ref{theorem2}, no need to calculate $p_1^{\ast}(\mathbf{x}_1)$ or $\mathbf{b}_1^{\ast}(\mathbf{x}_1)$ for each $\mathbf{x}_1$.
(vi) We also note that $\theta_1^{\ast}$ is not a function of $K_1$ or $\sigma_1^2$ (except that  $\theta_1^{\ast}$ not defined for $K_1=0$). This demonstrates that the optimal true location of the malicious vehicle does not depend on the inherent properties of the malicious channel (the channel between the malicious vehicle and the BS).
(vii) Following Theorem~\ref{theorem2}, we note that there is no \emph{unique} solution to the optimal true location  of the malicious vehicle  since \eqref{KL_minimum_xt} does not depend on $d_1$.
This is due to the fact that the malicious vehicle can adjust its transmit power to counteract the change of $d_1$ (i.e., $p_1^{\ast}(\mathbf{x}_1)$ is a function of $d_1$).

\begin{figure}[!t]
    \begin{center}
   {\includegraphics[width=3.5in, height=2.9in]{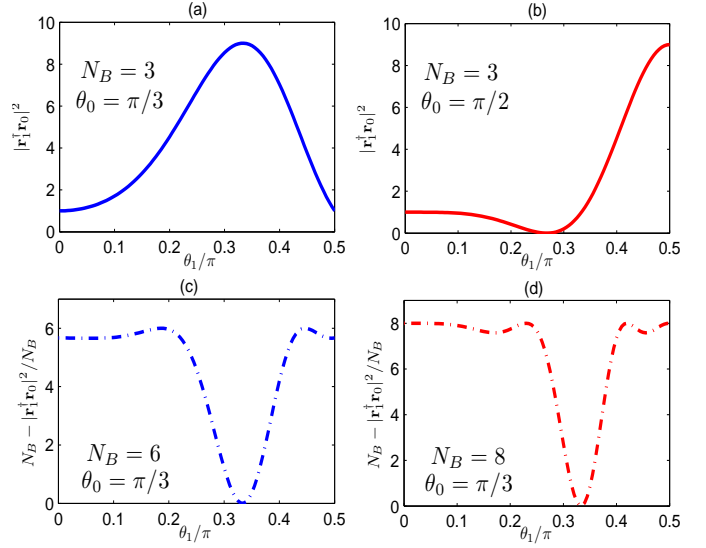}}
    \caption{$|\mathbf{r}_1^{\dag}\mathbf{r}_0|^2$ and $N_B - |\mathbf{r}_1^{\dag}\mathbf{r}_0|^2/N_B$ versus $\theta_1/\pi$ for different values of $N_B$ and $\theta_0$, where $\tau_B = \pi$.}\label{fig:norm}
    \end{center}
\end{figure}

Following \eqref{r_L_definition}, we have
\begin{align}\label{hb}
|\mathbf{r}_1^{\dag}\mathbf{r}_0|^2 =\left\{
\begin{array}{ll}
N_B^2, &  \cos \theta_0 = \cos \theta_1,\\
\left(\frac{\sin \left(\frac{1}{2}N_B \nu_{\theta} \right)}{\sin \left(\frac{1}{2} \nu_{\theta}\right)}\right)^2, & \cos \theta_0 \neq \cos \theta_1,
\end{array}
\right.
\end{align}
where $\nu_{\theta} = \tau_B (\cos \theta_0 - \cos \theta_1)$. To gain some further insights, we plot $|\mathbf{r}_1^{\dag}\mathbf{r}_0|^2$ and $N_B - |\mathbf{r}_1^{\dag}\mathbf{r}_0|^2/N_B$ versus $\theta_1/\pi$ in Fig.~\ref{fig:norm}. In Fig.~\ref{fig:norm}~(a), we first observe that the optimal attack is indeed at $\theta_1^{\ast} = \pm \theta_0$ (i.e., $\theta_1^{\ast} = \pm \theta_c$ due to $\theta_c = \theta_0$). Following \eqref{hb}, we note that the minimum KL divergence presented in \eqref{KL_minimum_xt} is zero for $\theta_1^{\ast} = \pm \theta_c$. This indicates that the malicious vehicle can perfectly imitate the signals expected from a legitimate vehicle at $\mathbf{x}_c$ if the malicious vehicle can set $\theta_1^{\ast} = \pm \theta_c$.\footnote{If additional BSs are in range of the claimed location this form of perfect attack can be neutralized. However, even in the one BS scenario (as we discuss later), when tracking is brought to bear on this issue this type of attack can minimized and even completely neutralized if constraints on the threat model are assumed (e.g., if the attacker is assumed to be another vehicle physically on the same highway as the legitimate vehicle).} In  Fig.~\ref{fig:norm}~(b) we also observe this effect, but this figure also illustrates that if  $\theta_1^{\ast} = \pm \theta_c$ was not possible (as was the case in this simulation in which the malicious vehicle could not access this angle due to the presence of a non-accessible area) then $|\mathbf{r}_1^{\dag}\mathbf{r}_0|^2$ does not necessarily increase as $\theta_1$ approaches $\theta_0$. This is due to the fact that $\theta_1$ minimizes $|\mathbf{r}_1^{\dag}\mathbf{r}_0|^2$ at
$\arccos \left(\cos \theta_0 + \frac{2 n_a \pi}{N_B \tau_a}\right)$ for $n_a = 1, \dots, N_B -1$. Comparing Fig.~\ref{fig:norm}~(c) with Fig.~\ref{fig:norm}~(d), we can see that $N_B - |\mathbf{r}_1^{\dag}\mathbf{r}_0|^2/N_B$ increases for the larger $N_B$ case.
This is consistent with the general rule that the minimum KL divergence presented in \eqref{KL_minimum_xt} increases as $N_B$ increases, and thus  indicates that the detection performance of the LVS increases as the number of antenna elements at the BS increases.

This above discussion also illustrates the very important role played by the constraint ${\|\mathbf{x}_c - \mathbf{x}_1\| \geq r_l}$ in  \eqref{opt_x1} in limiting any attack. For example, if a claimed location is within $r_l$ to the BS, and a building is between the claimed location and the malicious vehicle,  then no LOS component to the BS at the angle $\theta_1^{\ast} = \pm \theta_c$ is available to the malicious vehicle. Its actual optimal (now sub-optimal) attack location is then set at another angle. Assuming the malicious vehicle can always access a $\theta_1^{\ast} = \pm \theta_c$ location, with a non-zero LOS component to the BS, is therefore the most conservative scenario (worst-case scenario from the LVS perspective).

\subsection{Detection Performance of the LVS}

Without loss of generality, we first analyze the detection performance of the LVS based on any given $\theta_1$. Based on the proof of Theorem~\ref{theorem1}, we know that $\mathbf{R}_1 = \mathbf{R}_0$ when the malicious vehicle sets $p_1 = p_1^{\ast}(\mathbf{x}_1)$. Substituting \eqref{pdf_y0}, \eqref{opt_P1}, and \eqref{opt_b1} into \eqref{arbitrary}, the LRT decision rule presented in \eqref{arbitrary} can be written as
\begin{align}\label{decision_rule}
\mathbb{T}(\mathbf{y}) \begin{array}{c}
\overset{\Hoalt}{\geq} \\
\underset{\Honull}{<}
\end{array}%
\Gamma,
\end{align}
where $\mathbb{T}(\mathbf{y})$ is the test statistic given by
\begin{align}\label{test_statistic}
\mathbb{T}(\mathbf{y}) = 2 \text{Re}\{[\mathbf{m}_1^{\ast}(\theta_1) - \mathbf{m}_0]^{\dag}\mathbf{R}_0^{-1}\mathbf{y}\},
\end{align}
$\Gamma$ is the threshold for $\mathbb{T}(\mathbf{y})$ given by
\begin{align}\label{threshold_gamma}
\Gamma \!=\! \ln \lambda + \text{Re}\{[\mathbf{m}_1^{\ast}(\theta_1) \!-\! \mathbf{m}_0]^{\dag}\mathbf{R}_0^{-1}[\mathbf{m}_1^{\ast}(\theta_1) \!+\! \mathbf{m}_0]\},
\end{align}
and $\mathbf{m}_1^{\ast}(\theta_1)$ is obtained by substituting \eqref{opt_P1} and \eqref{opt_b1} into \eqref{mk_definition}, and is given by
\begin{align}
\mathbf{m}_1^{\ast}(\theta_1) = \sqrt{\frac{p_0 \mathrm{g}(d_0) K_0 N_0}{1+K_0}}\frac{\mathbf{r}_1 \mathbf{r}_1^{\dag} \mathbf{r}_0}{N_B}.
\end{align}
Next, we derive the false positive rate, $\alpha(\lambda,\theta_1)$, and the detection rate, $\beta(\lambda,\theta_1)$, of the LVS for any given $\theta_1$ in the following theorem.
\begin{theorem}\label{theorem3}
The false positive and detection rates of the LVS with $p_1 = p_1^{\ast}(\mathbf{x}_1)$ and $\mathbf{b}_1 = \mathbf{b}_1^{\ast}(\mathbf{x}_1)$ for any given $\theta_1$ are derived as
\begin{align}
\alpha(\lambda,\theta_1) &=\left\{
\begin{array}{ll}
\tilde{\alpha}(\lambda,\theta_1), &  \theta_1 \neq \pm \theta_c,\\
\mathbf{1}_A(-\Gamma) = \mathbf{1}_A(-\ln \lambda), & \theta_1 = \pm \theta_c,
\end{array}
\right. \label{alpha_r} \\
\beta(\lambda,\theta_1) &=\left\{
\begin{array}{ll}
\tilde{\beta}(\lambda,\theta_1), &  \theta_1 \neq \pm \theta_c,\\
\mathbf{1}_A(\Gamma) = \mathbf{1}_A(\ln \lambda), & \theta_1 = \pm \theta_c,
\end{array}
\right.\label{beta_r}
\end{align}
where
\begin{align}
\tilde{\alpha}(\lambda,\theta_1) &= \mathcal{Q}\left\{\frac{\Gamma - 2\text{Re}\{[\mathbf{m}_1^{\ast}(\theta_1) - \mathbf{m}_0]^{\dag}\mathbf{R}_0^{-1}\mathbf{m}_0\}}{\sqrt{2[\mathbf{m}_1^{\ast}(\theta_1) \!-\! \mathbf{m}_0]^{\dag}\mathbf{R}_0^{\!-\!1}[\mathbf{m}_1^{\ast}(\theta_1) \!-\! \mathbf{m}_0]}}\right\}\notag\\
&= \mathcal{Q}\left\{\frac{\ln \lambda + D(\theta_1)}{\sqrt{2 D(\theta_1)}}\right\},\label{alpha_rn}\\
\tilde{\beta}(\lambda,\theta_1) &= \mathcal{Q}\left\{\frac{\Gamma - 2\text{Re}\{[\mathbf{m}_1^{\ast}(\theta_1) - \mathbf{m}_0]^{\dag}\mathbf{R}_0^{-1}\mathbf{m}_1^{\ast}(\theta_1)\}}{\sqrt{2[\mathbf{m}_1^{\ast}(\theta_1) \!-\! \mathbf{m}_0]^{\dag}\mathbf{R}_0^{\!-\!1}[\mathbf{m}_1^{\ast}(\theta_1) \!-\! \mathbf{m}_0]}}\right\}\notag\\
&= \mathcal{Q}\left\{\frac{\ln \lambda - D(\theta_1)}{\sqrt{2 D(\theta_1)}}\right\},\label{beta_rn}
\end{align}
$\mathcal{Q}(x) = \frac{1}{\sqrt{2 \pi}}\int_x^{\infty}\exp\left(-\frac{t^2}{2}\right)d t$, $D(\theta_1)$ is the minimum KL divergence for any $\theta_1$ given by (following \eqref{KL_minimum_xt})
\begin{align}
D(\theta_1) &= [\mathbf{m}_1^{\ast}(\theta_1) \!-\! \mathbf{m}_0]^{\dag}\mathbf{R}_0^{\!-\!1}[\mathbf{m}_1^{\ast}(\theta_1) \!-\! \mathbf{m}_0]\notag \\
&= \frac{p_0 g(d_0)K_0 N_0}{p_0 g(d_0) + \sigma_0^2 (1+K_0)} \left( N_B - \frac{|\mathbf{r}_1^{\dag}\mathbf{r}_0|^2}{N_B}\right),
\end{align}
and $\mathbf{1}_A(x)$ is a indicator function defined by
\begin{align}
\mathbf{1}_A(x) =\left\{
\begin{array}{ll}
1, &  x \geq 0,\\
0, & x < 0.
\end{array}
\right.
\end{align}
\end{theorem}
\begin{IEEEproof}
Following \eqref{test_statistic}, we derive the distributions of the test statistic $\mathbb{T}(\mathbf{y})$ for $\theta_1 \neq \pm \theta_c$ under $\Hnull$ and $\Halt$ as follows
\begin{align}
\mathbb{T}(\mathbf{y})|\Hnull &\sim \mathcal{N}\bigg(2\text{Re}\{[\mathbf{m}_1(\theta_1) \!-\! \mathbf{m}_0]^{\dag}\mathbf{R}_0^{\!-\!1}\mathbf{m}_0\}, \bigg.\notag \\
&~~~~~~\bigg.2[\mathbf{m}_1^{\ast}(\theta_1) \!-\! \mathbf{m}_0]^{\dag}\mathbf{R}_0^{\!-\!1}[\mathbf{m}_1^{\ast}(\theta_1) \!-\! \mathbf{m}_0]\bigg),\label{d_t0}\\
\mathbb{T}(\mathbf{y})|\Halt &\sim \mathcal{N}\bigg(2\text{Re}\{[\mathbf{m}_1(\theta_1) \!-\! \mathbf{m}_0]^{\dag}\mathbf{R}_0^{\!-\!1}\mathbf{m}_1(\theta_1)\}, \bigg.\notag \\
&~~~~~~\bigg.2[\mathbf{m}_1^{\ast}(\theta_1) \!-\! \mathbf{m}_0]^{\dag}\mathbf{R}_0^{\!-\!1}[\mathbf{m}_1^{\ast}(\theta_1) \!-\! \mathbf{m}_0]\bigg).\label{d_t1}
\end{align}
Based on the decision rule in \eqref{decision_rule} and the definitions of the false positive and detection rates, we obtain the false positive and detection rates for $\theta_1 \neq \pm \theta_c$ in \eqref{alpha_rn} and \eqref{beta_rn} after some algebraic manipulations. For $\theta_1 = \pm \theta_c$, following \eqref{test_statistic} and \eqref{threshold_gamma} we have $\mathbb{T}(\mathbf{y}) = 0$ and $\Gamma = \ln \lambda$. Then, based on the decision rule presented in \eqref{decision_rule} we obtain the desirable results for $\theta_1 = \pm \theta_c$ as detailed in \eqref{alpha_r} and \eqref{beta_r}. This completes the proof of Theorem~\ref{theorem3}.
\end{IEEEproof}

We note that both $\alpha(\lambda,\theta_1)$ and $\beta(\lambda,\theta_1)$ are functions of $D(\theta_1)$, which is the minimum KL divergence for a given $\theta_1$ presented in \eqref{KL_minimum_xt}. Based on the properties of $\mathcal{Q}(x)$ and the expressions for $\alpha(\lambda,\theta_1)$ and $\beta(\lambda,\theta_1)$, we know that the detection performance of the LVS increases as $D(\theta_1)$ increases (e.g., for a given $\alpha(\lambda,\theta_1)$, $\beta(\lambda,\theta_1)$ increases as $D(\theta_1)$ increases).
This confirms that the malicious vehicle is to search for $\mathbf{x}_1^{\ast}$ through minimizing the minimum KL divergence presented in \eqref{KL_minimum_xt}.
By setting $\lambda = \lambda^{\ast}$, following \eqref{total_error} the minimum total error conditioned on a $\theta_1$ can be expressed as \cite{barkat2005signal}
\begin{align}
\epsilon^{\ast} (\theta_1) = P_0 \alpha(\lambda^{\ast},\theta_1) + (1-P_0)\left(1 - \beta(\lambda^{\ast},\theta_1)\right).
\end{align}
We note that the detection performance of the LVS based on the malicious vehicle's optimal true location can be obtained by substituting $\theta_1^{\ast}$ into our derived $\alpha(\lambda,\theta_1)$ and $\beta(\lambda,\theta_1)$.

In summarizing this section we note the following. The analysis given above provides the optimal decision framework for the LVS, and provides analytical solutions for the detection performance of the LVS as a function of the Rician $K$-factors. The performance analysis given is valid for scenarios in which the optimal $\theta_1$ value is forbidden to the malicious vehicle. The closest work to the results of this section are perhaps those detailed in \cite{yan2014location}. However, in \cite{yan2014location},  due to the assumption of a severely restrictive threat model in which the malicious vehicle must be on the same road as the legitimate vehicle, the allowed attack scenarios are severely constrained and far from optimal (e.g., $\theta_1$'s allowed are small). Also, for simplicity, the beamformer in \cite{yan2014location} is set using a single parameter, which means it does not possess independently-set  antenna coefficients. Such a simple beamformer provides poorer performance (relative to the solutions presented here) when an attack from a non-optimal location is launched.

\section{Location Verification System \\ with Tracking}\label{sec_LVS-S}

In this section we examine the LVS when tracking information on the claimed location is available. That is,  when claimed locations and BS measurements are available at multiple (sequential) time slots ($T \geq 2$). We refer to this LVS as the \emph{tracking LVS}. We first present the decision rule adopted in this tracking LVS, and then present the optimal attack strategy of the malicious vehicle against the tracking LVS. Finally, we analyze the detection performance of the tracking LVS based on this optimal attack strategy.

\subsection{Decision Rule of the Tracking LVS}

In the tracking LVS we assume that we collect one $\mathbf{y}(t)$ for each claimed location $\mathbf{x}_c(t)$. There are several questions we could pose given the introduction of tracking information to the LVS. However, perhaps the most pragmatic question for a tracking LVS is how to make an optimal decision (e.g., minimize the total error) on whether the vehicle is legitimate or malicious given the  last sequence of observations at its disposal. An important system-model issue in tracking mode is that we will let the tracking LVS randomly select the number of time slots to be used prior to each LVS decision. That is, once a decision is made based on say $T_a$ time slots, the next decision is made independently based on the next say $T_b$ time slots, where  $T_a$ and $T_b $ are specific realizations of the random variable $T$.
Operationally, this means the specific realizations of $T$ will always be unknown to the malicious vehicle.\footnote{The (non-tracking) LVS  discussed earlier is now seen as  a special case of the tracking LVS with the realization of $T$ always set equal to one and without the additional constraint $r_u$. Note, due to the additional constraint $r_u$, the tracking solution in general is not identical to a solution derived from the direct use of individual unit ($T=1$) timeslot decisions. } Henceforth, when we use $T$ we will mean a realization of the random variable $T$.
Under such conditions the optimal decision rule (per decision) for the tracking LVS will be an expanded version of our previously utilized LRT, namely,
\begin{equation}\label{arbitrary_t}
\Lambda_{track}\left(\mathbf{Y}(T)\right)
\begin{array}{c}
\overset{\Hoalt}{\geq} \\
\underset{\Honull}{<}
\end{array}%
\lambda_{track},
\end{equation}
where $\mathbf{Y}(T) = [\mathbf{y}(1), \cdots, \mathbf{y}(T)]$, $\Lambda_{track}\left(\mathbf{Y}(T)\right)$ is the likelihood ratio of $\mathbf{Y}(T)$ given by
\begin{align}
\Lambda_{track}\left(\mathbf{Y}(T)\right) = \frac{f\left(\mathbf{Y}(T)|\mathbf{p}_1(T), \mathbf{B}_1(T), \mathbf{X}_1(T), \Halt\right)}{f\left(\mathbf{Y}(T)|\Hnull\right)},
\end{align}
$\mathbf{p}_1(T) = [p_1(1), \cdots, p_1(T)]$, $\mathbf{B}_1(T) = [\mathbf{b}_1(1), \cdots, \mathbf{b}_1(T)]$, $\mathbf{X}_1(T) = [\mathbf{x}_1(1), \cdots, \mathbf{x}_1(T)]$, and $\lambda_{track}$ is the threshold for $\Lambda_{track}\left(\mathbf{Y}(T)\right)$.
Since $\mathbf{y}(t)$ are independent for different $t$, we further have
\begin{align}\label{likelihood_ratio_t}
\Lambda_{track}\left(\mathbf{Y}(T)\right) = \frac{\prod_{t = 1}^T f\left(\mathbf{y}(t)|p_1(t), \mathbf{b}_1(t), \mathbf{x}_1(t), \Halt\right)}{\prod_{t = 1}^T f\left(\mathbf{y}(t)|\Hnull\right)}.
\end{align}
Again, the false positive rate of the tracking LVS is given by $\alpha_{track}(\lambda_{track}) = \Pr\left(\Lambda_{track}\left(\mathbf{Y}(T)\right) > \lambda_{track}|\Hnull\right)$, and the detection rate of the tracking LVS is given by $\beta_{track}(\lambda_{track}) =\Pr\left(\Lambda_{track}\left(\mathbf{Y}(T)\right) > \lambda_{track}|\Halt\right)$.
The specific value of $\lambda_{track}$ can be set based on a  methodology similar to that used in setting $\lambda$. We again adopt the total error as the unique performance metric to evaluate the tracking LVS. The optimal value of $\lambda_{track}$ that minimizes the total error of the tracking LVS is given by $\lambda_{track}^{\ast} = P_0/(1-P_0)$ \cite{barkat2005signal}.

\subsection{Optimal Attack Strategy Against the Tracking LVS}

Knowing \eqref{arbitrary_t}, in order to minimize the detection rate, the malicious vehicle is to minimize the following KL divergence \cite{eguchi2006interpreting}
\begin{align}
&D_{KL}\left(f\left(\mathbf{Y}(T)|\mathbf{p}_1(T), \mathbf{B}_1(T), \mathbf{X}_1(T), \Halt\right)|| \right. \notag  \\
&\qquad \quad \left.
f\left(\mathbf{Y}(T)|\Hnull\right)\right) \notag \\
&= \int \left[ \ln \Lambda_{track} (\mathbf{Y}(T)) \right] \times \notag \\
& \qquad f\left(\mathbf{Y}(T)|\mathbf{p}_1(T), \mathbf{B}_1(T), \mathbf{X}_1(T), \Halt\right) d \mathbf{Y}(T) \notag\\
&= \int \left[ \sum_{t = 1}^T \ln \Lambda (\mathbf{y}(t)) \right] \prod_{t = 1}^T f\left(\mathbf{y}(t)|p_1(t), \mathbf{b}_1(t), \mathbf{x}_1(t), \Halt\right)\notag \\
& \qquad \quad \times  d \mathbf{Y}(T) \notag
\end{align}
\begin{align}\label{KL_definition_t}
&= \sum_{t = 1}^T D_{KL}\left(f\left(\mathbf{y}(t)|p_1(t), \mathbf{b}_1(t), \mathbf{x}_1(t), \Halt\right)||f\left(\mathbf{y}(t)|\Hnull\right)\right).
\end{align}
Based on \eqref{KL_definition_t}, we know that the KL divergence for $t = 1, 2, \cdots, T$ is the sum of the KL divergence presented in \eqref{KL_definition} for each $t$. We also can see that the KL divergence at $t$ is independent of the system settings at other time slots. This indicates that the malicious vehicle can optimize all the parameters under his control at $t$ (e.g., $p_1(t)$, $\mathbf{b}_1(t)$, and $\mathbf{x}_1(t)$) by considering only the system settings for the current time slot $t$ (e.g., the values of $\mathbf{x}_c(t)$, $\sigma_0^2(t)$, and $\sigma_1^2(t)$).
As such, the optimal attack strategy for the malicious vehicle is to optimize all parameters under its control for the current time slot. To this end, for each $t$ the malicious vehicle first optimizes $p_1(t)$ and $\mathbf{b}_1(t)$ according to Theorem~\ref{theorem1} for any given $\mathbf{x}_1(t)$. Then, the malicious vehicle is to optimize $\mathbf{x}_1(t)$ under some constraints detailed in the following.
For $\mathbf{x}_c(1)$, the malicious vehicle can optimize $\mathbf{x}_1(t)$ according to Theorem~\ref{theorem2}. We would like to highlight that in addition to $|\mathbf{x}_c(t) - \mathbf{x}_1(t)| \geq r_l$ there is another constraint on $\mathbf{x}_1(t)$ for $t \geq 2$, which is that $|\mathbf{x}_1^{\ast}(t-1) -\mathbf{x}_1(t)| \leq r_u$, where $r_u$ can be determined through imposition of a realistic vehicle speed limitation. This is due to the fact that the malicious vehicle cannot move too far away from its previous location (i.e., its location in the previous  time slot). Then, the optimal $\theta_1(t)$ for $t \geq 2$  is given by
\begin{align}\label{x1_sequential}
\theta_1^{\ast}(t) = \argmax_{\|\mathbf{x}_c(t) - \mathbf{x}_1(t)\| \geq r_l, \atop \|\mathbf{x}_1^{\ast}(t-1) -\mathbf{x}_1(t)\| \leq r_u} |\mathbf{r}_1^{\dag}(t)\mathbf{r}_0(t)|^2.
\end{align}
We note that the optimal attack strategy  against the tracking LVS for the malicious vehicle is to find an angle $\theta_1^{\ast}(t) = \pm \theta_c(t)$ with a non-zero LOS component towards the BS for every time slot. Should the two distance constraints imposed on the malicious vehicle make $\theta_1^{\ast}(t) = \pm \theta_c(t)$ impossible, then a sub-optimal attack at $\theta_1^{\ast}(t) \neq \pm \theta_c(t)$ must take place at some of the time slots.

\subsection{Detection Performance of the Tracking LVS}

Without loss of generality, we analyze the detection performance of the tracking LVS for any given $\bm{\theta}_1(T) = [\theta_1(1), \cdots, \theta_1(T)]$ by considering $p_1(t) = p_1^{\ast}(\mathbf{x}_1(t))$ and $\mathbf{b}_1(t) = \mathbf{b}_1^{\ast}(\mathbf{x}_1(t))$. We denote the track of claimed locations as $\bm{\theta}_c(T) = [\theta_c(1), \cdots, \theta_c(T)]$. Following \eqref{likelihood_ratio_t}, the LRT decision rule presented in \eqref{arbitrary_t} can be rewritten as
\begin{align}\label{decision_rule_track}
\mathbb{T}_{track}(\mathbf{Y}(T)) \begin{array}{c}
\overset{\Hoalt}{\geq} \\
\underset{\Honull}{<}
\end{array}%
\Gamma_{track},
\end{align}
where $\mathbb{T}_{track}(\mathbf{Y}(T))$ is the test statistic given by
\begin{align}\label{test_statistic_track}
\mathbb{T}_{track}(\mathbf{Y}(T)) = 2 \text{Re}\left\{\sum_{t=1}^T[\mathbf{m}_1^{\ast}(\theta_1(t)) - \mathbf{m}_0(t)]^{\dag}\mathbf{R}_0^{-1}\mathbf{y}(t)\right\},
\end{align}
and $\Gamma_{track}$ is the threshold for $\mathbb{T}_{track}(\mathbf{Y}(T))$ given by
\begin{align}
&\Gamma_{track} \!=\! \ln \lambda_{track} + \notag \\
&\text{Re}\left\{\sum_{t=1}^T [\mathbf{m}_1^{\ast}(\theta_1(t)) \!-\! \mathbf{m}_0(t)]^{\dag}\mathbf{R}_0^{-1}[\mathbf{m}_1^{\ast}(\theta_1(t)) \!+\! \mathbf{m}_0(t)]\right\}.
\end{align}
We then derive the false positive rate, $\alpha_{track}(\lambda_{track},\bm{\theta}_1(T))$, and the detection rate, $\beta_{track}(\lambda_{track},\bm{\theta}_1(T))$, of the tracking LVS for any given $\bm{\theta}_1(T)$ in the following theorem.
\begin{theorem}\label{theorem4}
The false positive rate and the detection rate of the tracking LVS for any given $\bm{\theta}_1(T)$ are derived as
\begin{align}
&\alpha_{track}(\lambda_{track},\bm{\theta}_1(T)) \notag \\
&\quad =\left\{
\begin{array}{ll}
\tilde{\alpha}_{track}(\lambda_{track},\bm{\theta}_1(T)), &  \bm{\theta}_1(T) \neq \pm \bm{\theta}_c(T),\\
\mathbf{1}_A(-\ln \lambda_{track}), & \bm{\theta}_1(T) = \pm \bm{\theta}_c(T),
\end{array}
\right. \label{alpha_rt} \\
&\beta_{track}(\lambda_{track},\bm{\theta}_1(T)) \notag \\
&\quad =\left\{
\begin{array}{ll}
\tilde{\beta}_{track}(\lambda_{track},\bm{\theta}_1(T)), &  \bm{\theta}_1(T) \neq \pm \bm{\theta}_c(T),\\
\mathbf{1}_A(\ln \lambda_{track}), & \bm{\theta}_1(T) = \pm \bm{\theta}_c(T),
\end{array}
\right.\label{beta_rt}
\end{align}
where
\begin{align}
\tilde{\alpha}_{track}(\lambda_{track},\bm{\theta}_1(T))
&\!=\! \mathcal{Q}\left\{\frac{\ln \lambda_{track} + D_{track}(\bm{\theta}_1(T))}{\sqrt{2 D_{track}(\bm{\theta}_1(T))}}\right\},\label{alpha_rtn}\\
\tilde{\beta}_{track}(\lambda_{track},\bm{\theta}_1(T))
&\!=\! \mathcal{Q}\left\{\frac{\ln \lambda_{track} - D_{track}(\bm{\theta}_1(T))}{\sqrt{2 D_{track}(\bm{\theta}_1(T))}}\right\},\label{beta_rtn}
\end{align}
and $D_{track}(\bm{\theta}_1(T))$ is the minimum KL divergence for any given $\bm{\theta}_1(T)$, which is given by (following \eqref{KL_minimum_xt} and \eqref{KL_definition_t})
\begin{align}\label{KL_minimum_xt_tracking}
&D_{track}(\bm{\theta}_1(T)) \notag \\
&=\sum_{t = 1}^T D_{KL}\left(f\left(\mathbf{y}(t)|p_1^{\ast}(t), \mathbf{b}_1^{\ast}(t), \mathbf{x}_1(t), \Halt\right)||f\left(\mathbf{y}(t)|\Hnull\right)\right)\notag \\
&=\sum_{t=1}^T \frac{p_0 g(d_0(t))K_0(t) N_0}{p_0 g(d_0(t)) \!+\! \sigma_0^2(t) (1\!+\!K_0(t))} \left( N_B \!-\! \frac{|\mathbf{r}_1(t)^{\dag}\mathbf{r}_0(t)|^2}{N_B}\right).
\end{align}
\end{theorem}
\begin{IEEEproof}
The proof of Theorem~\ref{theorem4} is very similar to that of Theorem~\ref{theorem3}, we therefore omit it here.
\end{IEEEproof}

The minimum total error of the tracking LVS for any given $\bm{\theta}_1(T)$ is \cite{barkat2005signal}
\begin{align}
\epsilon_{track}^{\ast} (\bm{\theta}_1(T)) &= P_0 \alpha_{track}(\lambda_{track}^{\ast},\bm{\theta}_1(T))\notag \\
&\quad + (1-P_0) \left(1 - \beta_{track}(\lambda_{track}^{\ast},\bm{\theta}_1(T))\right).
\end{align}
We note that the minimum KL divergence provided in \eqref{KL_minimum_xt} is greater than zero for any $\mathbf{x}_1(t)$ as long as $\theta_1(t) \neq \pm \theta_c(t)$. As such, $D_{track}(\bm{\theta}_1(T))$ monotonically increases as $T$ increases for $\theta_1(t) \neq \pm \theta_c(t)$. This demonstrates that the detection performance of the tracking LVS increases as $T$ increases as long as $\theta_1(t) \neq \pm \theta_c(t)$ (e.g., $\epsilon_{track}^{\ast} (\bm{\theta}_1(T))$ decreases as $T$ increases).

In summarizing this section we note the following. The above analysis on the tracking LVS makes the following key points. Under the assumption that $T$ is randomly selected \emph{per decision} by the tracking LVS, the optimal decision framework is a reasonably extension of the non-tracking framework. The optimal attack scenario is for the malicious vehicle to be at $\theta_1^{\ast}(t) = \pm \theta_c(t)$. However, physical constraints (such as limited speed) may make this impossible. The next sub-optimal malicious vehicle location can then be calculated - and this location may not necessarily be the $\theta_1(t)$ closest to $\theta_0(t)$ with non-zero LOS components.
The performance of the tracking LVS under any potential sequence of the malicious vehicle's locations is provided analytically.
The closest work to the analysis presented in this section is perhaps \cite{malaney2007wireless}, in which  a wireless intrusion detection system based on the utilization of position tracking and the localization error bounds of Extended Kalman Filters is developed. It is shown in \cite{malaney2007wireless} that the detection errors of the system with tracking information can be an order of magnitude smaller relative to that of the system with only a static location. However, the optimality of the location spoofing detection system with tracking was not discussed in \cite{malaney2007wireless}.

\section{Numerical Results}\label{sec_numerical}

In this section, we present numerical simulations to verify the accuracy of our provided analysis on the LVS and the tracking LVS. We also provide some useful insights on the impact of $p_0$, $\theta_1^{\ast}$, $N_B$, $N_0$, and $K_0$ on the detection performance of the LVS. We further examine the impact of $K_1$ and $\sigma_1^2$ on $N_1^{\ast}$.

\subsection{Numerical Results for the LVS}\label{sec_numericala}

\begin{figure}[!t]
    \begin{center}
   {\includegraphics[width=3.5in, height=2.9in]{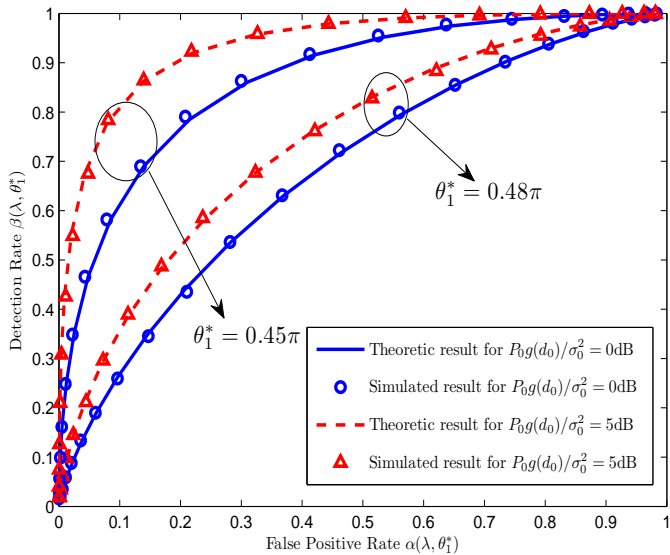}}
    \caption{ROC curves of the LVS for $N_B = 4$, $N_0=3$, $N_1 \geq N_1^{\ast}$, $\theta_0 = \pi/2$, $K_0 = 1$dB, $\sigma_0^2 = \sigma_1^2 = 0$dB, $p_1 = p_1^{\ast}(\mathbf{x}_1)$, and $\mathbf{b}_1 = \mathbf{b}_1^{\ast}(\mathbf{x}_1)$.}\label{fig:fig1}
    \end{center}
\end{figure}

We first consider the LVS (i.e., the non-tracking LVS) and thus we drop the index $(t)$ in this subsection.
In Fig.~\ref{fig:fig1}, we present the Receiver Operating Characteristic (ROC) curve of the LVS.
In this figure, we first observe that the Monte Carlo simulations precisely match the theoretic results, which confirms our analysis presented in Theorem~\ref{theorem3}. We also observe that the ROC curves for $p_0 g(d_0)/\sigma_0^2 = 5$dB dominate the ROC curves for $p_0 g(d_0)/\sigma_0^2 = 0$dB. This observation demonstrates that the detection performance of the LVS increases as the signal-to-noise ratio (SNR) of the legitimate channel (the channel between the BS and the legitimate vehicle) increases. As expected, we further observe that the ROC curve shifts towards the right-lower corner as $\theta_1^{\ast}$ moves closer to $\theta_0$.

\begin{figure}[!t]
    \begin{center}
   {\includegraphics[width=3.5in, height=2.9in]{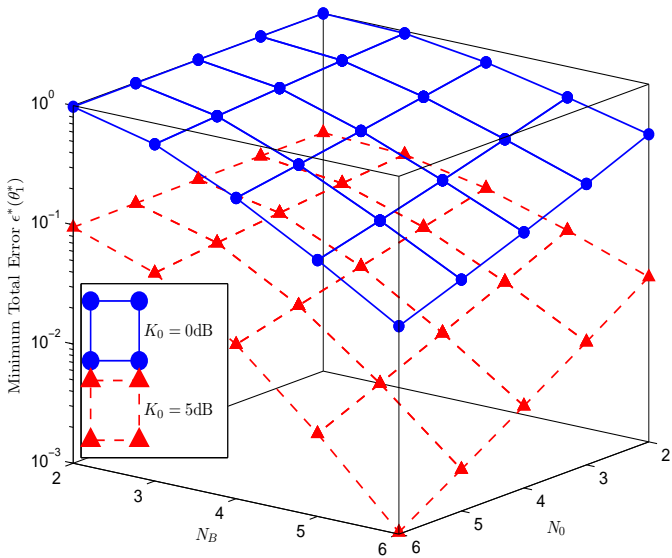}}
    \caption{Minimum total error of the LVS versus $N_B$ and $N_0$ for $P_0 = 0.9$, $\theta_0 = \pi/3$, $\theta_1^{\ast} = \pi/4$, $K_1 = 0$dB, $p_0 g(d_0)/\sigma_0^2 = 0$dB, $p_1 = p_1^{\ast}(\mathbf{x}_1)$, $\mathbf{b}_1 = \mathbf{b}_1^{\ast}(\mathbf{x}_1)$, and $N_1 \geq N_1^{\ast}$.}\label{fig:fig2}
    \end{center}
\end{figure}

\begin{figure}[!t]
    \begin{center}
   {\includegraphics[width=3.5in, height=2.9in]{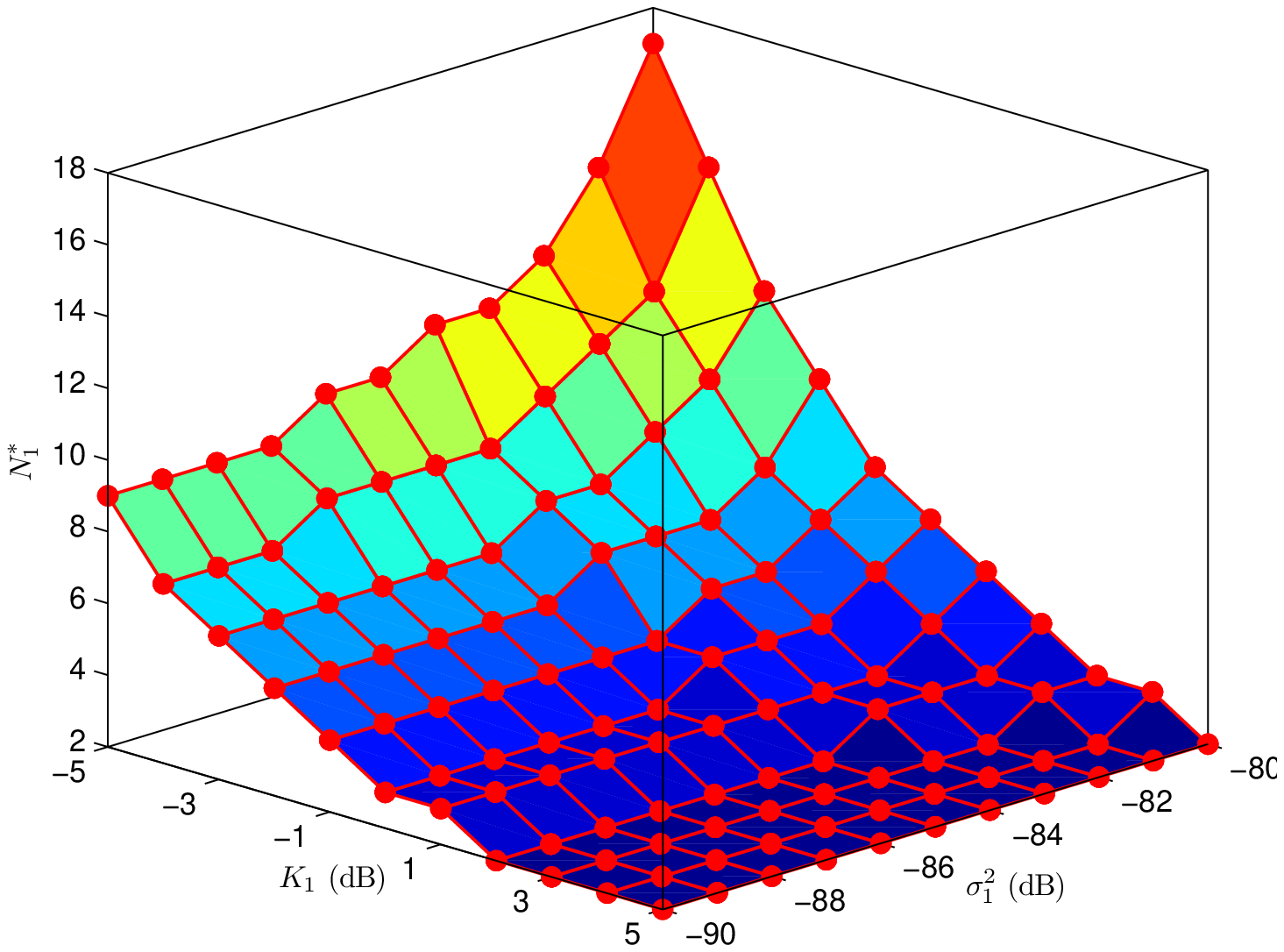}}
    \caption{$N_1^{\ast}$ versus $K_1$ and $\sigma_1^2$ for $N_0 = 3$, $p_0 g(d_0) = -75$dB, $\sigma_0^2 = -85$dB, and $K_0 = 0$dB.}\label{fig:fig3}
    \end{center}
\end{figure}

In Fig.~\ref{fig:fig2}, we present the minimum total error $\epsilon(\theta_1^{\ast})$ versus the number of antenna elements at the legitimate vehicle ($N_0$) and the number of antenna elements at the BS ($N_B$). As expected, we first observe that $\epsilon(\theta_1^{\ast})$ decreases as $N_B$ or $N_0$ increases.
We also observe that $\epsilon(\theta_1^{\ast})$ decreases as the Rician $K$-factor of the legitimate channel ($K_0$) increases.
From the simulations to obtain Fig.~\ref{fig:fig2}, we confirm $N_1^{\ast}$ increases as $N_0$ or $K_0$ increases, but is not a function of $N_B$.

In Fig.~\ref{fig:fig3}, we plot $N_1^{\ast}$ versus Rician $K$-factor of the malicious channel, $K_1$, and the noise variance of the malicious channel, $\sigma_1^2$. As expected from \eqref{opt_N1}, we first observe that $N_1^{\ast}$ increases as $K_1$ decreases or $\sigma_1^2$ increases. This demonstrates that $N_1^{\ast}$ is highly dependent on the inherent properties of the malicious channel.
We also observe that $N_1^{\ast}$ is  a reasonable value (e.g., $15$) even when $K_1$ is  small (e.g., $-5$dB).

\subsection{Numerical Results for the Tracking LVS}\label{sec_numericalb}

\begin{figure}[!t]
    \begin{center}
   {\includegraphics[width=3.5in, height=2.9in]{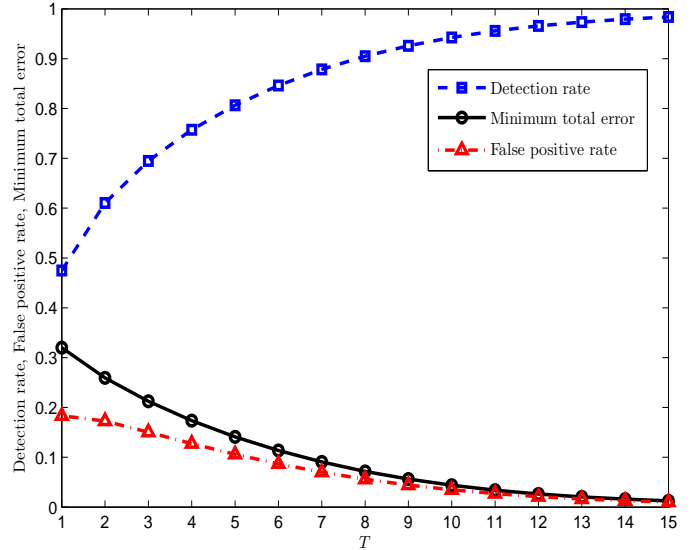}}
    \caption{False positive rate, detection rate, and minimum total error of the tracking LVS versus $T$ for $P_0 = 0.6$, $N_B = 3$, $N_0=2$, $N_1 \geq N_1^{\ast}$, $p_0 =30$dB, $K_0 = -10$dB, $\xi = 3$, $c = 3 \times 10^8$m/s, $f_0 = 5.9$GHz, $r_l = 100$m, $r_u = 3$m, $\tau_B = \pi$, and $\mathbf{x}_1(t) = \mathbf{x}_1^{\ast}(t)$.}\label{fig:LVS_T}
    \end{center}
\end{figure}

In Fig.~\ref{fig:LVS_T}, we examine the impact of $T$ on the detection performance of the tracking LVS.
In the simulations to obtain Fig.~\ref{fig:LVS_T}, we have assumed the claimed location $\mathbf{x}_0(t)$ is moving towards the BS along a straight line with a constant velocity 20km/h and $\mathbf{x}_0(1) = [10 \sqrt{2}, \pi/4]$. We have also assumed that $\mathbf{x}_1(t)$ is on the straight line and $K_0$ is a constant for all $\mathbf{x}_0(t)$. These settings mimic a practical VANETs scenario, where the BS is on the roadside, the legitimate vehicle is moving along the road towards the BS, and the malicious vehicle is also on the same road. The observation frequency and claimed-location reception are both set at 10 Hz (10 time slots per second). Other parameters adopted are specified in the caption of Fig.~\ref{fig:LVS_T}. As expected, we observe that the false positive rate and the minimum total error decreases as $T$ increases and the detection rate increases as $T$ increases. With the aid of the derived false positive and detection rates provided in \eqref{alpha_rt} and \eqref{beta_rt}, we can quantify the detection performance improvement brought by increased $T$. For example, the minimum total error for $T = 10$ is only about $30\%$ of that for $T=1$.

Finally, in this section we note the effect some of our channel and system model assumptions have on our results. More specifically, we probe circumstances where non-zero errors on the claimed location are present (inclusion of location errors also probes the impact of other issues such as inaccuracies in the $K$ map and potential shadowing effects). In general, we find such real-world effects have a limited impact on our results. For example, for the localization error (average distance between the estimated positions and the real positions) of 5 meters we find the results of Fig.~\ref{fig:LVS_T} are impacted only at the 10\% level (e.g., the false positive rate for this localization error is about 10\% higher than that for the zero localization error).

\section{Discussion}\label{sec_discussion}

\subsection{Other Antenna Arrays}

Although we assumed that the malicious vehicle is equipped with a ULA, our main analysis provided in this work still holds if the ULA is replaced by other antenna arrays (e.g., non-uniform linear arrays, circular arrays, rectangle arrays). If the malicious vehicle is equipped with other antenna arrays, only \eqref{t_L_definition} under $\Halt$ will be modified. For example, if the ULA at the malicious vehicle in Fig.~\ref{fig:system} is replaced by a Uniform Circular Array (UCA) centered at the malicious vehicle, \eqref{t_L_definition} under $\Halt$ will be replaced by the following equation (dropping the index $t$) \cite{ioannides2005uniform}
\begin{align}
\mathbf{t}_1  &= \left[\exp(-j \tau_1^c \cos  \phi_1),\!\cdots\!,\exp(-j \tau_1^c \cos  \phi_{N_1})\right], \label{t_L_uca}
\end{align}
where $\tau_1^c = 2 \pi f_c a_1 /c$, $a_1$ is the radius of the UCA at the malicious vehicle, and $\phi_m = 2 \pi (m-1)/{N_1} + \phi_1$ (where $m = 1, 2, \cdots, N_1$) is the angle measured counterclockwise from the reference line (the line connecting the center of the malicious vehicle's UCA and the center of the ULA at the BS) to the $m$-th antenna element of the UCA.
Based on \eqref{t_L_uca} and $\overline{\mathbf{H}}_1 = \mathbf{r}_{1}\mathbf{t}_1$ we can see that $\overline{\mathbf{H}}_1$ still only contains the directional information of the malicious channel (i.e., $\overline{\mathbf{H}}_1$ only depends on $\theta_1$ and $\phi_1$). Noting $\mathbf{Q} \propto \overline{\mathbf{H}}_1^{\dag} \overline{\mathbf{H}}_1$, we know that the matrix $\mathbf{Q}$ involved in Theorem~\ref{theorem1} is still a rank-1 matrix due to $\mathbf{r}_{k}^{\dag}\mathbf{r}_{k} = N_B$. In addition, as we have shown in Theorem~\ref{theorem3} the detection performance of the LVS is not a function of $\mathbf{t}_1$ as long as the malicious vehicle adopts the optimal transmit power and beamformer. As such, all the analysis provided earlier still holds exactly for the case where the malicious vehicle is equipped with the UCA. That is, the use of a UCA provides the attacker no additional benefit. Finally, we note our analysis can be readily adapted to cases where antenna arrays under the control of the LVS (e.g., at the BS and legitimate vehicle) are also non-linear arrays.


\subsection{Colluding Attacks}

We note that in practice the malicious vehicle may launch colluding attacks to the LVS and the tracking LVS by cooperating with other malicious vehicles.
However,  colluding attacks of any form cannot bring any additional benefits to the malicious vehicle that can set $\theta_1^{\ast}(t) = \pm \theta_c(t)$ at every decision step. This is because the minimum KL divergence presented in \eqref{KL_minimum_xt} will always be zero when $\theta_1^{\ast}(t) = \pm \theta_c(t)$. This is the case for both the (non-tracking) LVS  and the tracking LVS.

Considering the case where $\theta_1^{\ast}(t) \neq \pm \theta_c(t)$, there are two general specific attack strategies that can adopted by the colluding malicious vehicles, single-transmission attacks and multiple-transmission attacks.
In the single-transmission attack only one of the colluding malicious vehicles is active and transmitting signals. As such, the collusion in this type of attack takes the form of information-sharing and the subsequent decision of which vehicle is in the optimal location to launch an attack. The single-transmission attack can help a malicious vehicle against the tracking LVS (but not the non-tracking LVS). This is because the colluding malicious vehicles can potentially cooperatively select their true locations over different time slots in order to avoid the second constraint in \eqref{x1_sequential}, i.e. $\|\mathbf{x}_1^{\ast}(t-1) -\mathbf{x}_1(t)\| \leq r_u$. As the number of colluding malicious vehicles approach infinity, this constraint can be removed from \eqref{x1_sequential} entirely.
In the multiple-transmission attack, all the colluding malicious vehicles are active and transmitting signals simultaneously. As such, the collusion  takes the form of information-sharing and the subsequent decisions on the optimal transmit power, beamformer, and locations of the colluding malicious vehicles. Obviously such a sophisticated attack could outperform the single transmission attack in the general scenario.
But again we stress that when  $\theta_1^{\ast}(t) = \pm \theta_c(t)$ is allowed none of these colluding attacks are of importance. As such, adopting the detection rates for $\theta_1^{\ast}(t) = \pm \theta_c(t)$ always provides a worst-case bound for the LVS and the tracking LVS.

\section{Conclusion}\label{sec_conclusion}

In this work we have proposed a generic LVS framework for multi-antenna communication systems, and conducted a detailed  analysis of the framework's location authentication performance.
Although our work is general and can cover many  application scenarios, we have focussed here on the emerging VANETs paradigm under the assumption of Rician channels.
Such channels are anticipated to dominate real-world VANETs communication conditions. The LVS solution we have proposed is very general and  provides a foundation for all optimal location authentication schemes in the VANET scenario. Taking as inputs a claimed location and raw observations across the receiving BS antennas, our LVS checks its knowledge of the Rician channel conditions in its vicinity, forms a view as to the optimal attack location (from the attacker's viewpoint), and then outputs a binary decision on whether a vehicle is providing a legitimate location.
Our analysis quantifies the dependence between the detection performance limit of the LVS and the Rician $K$-factor of the legitimate channel, and formally reveals that the LVS performance limit is independent of the properties of the malicious channel.
In addition, our analysis discloses  that once the malicious vehicle's number of antennas reaches a derived bound, further increases in this number does not reduce the detection rate. We also formalized the optimal decision rule when  tracking information is added to the LVS.
The work presented here will be of importance to  emerging intelligent vehicular network scenarios, particularly in relation to certificate revocation schemes within IEEE 1609.2.


\begin{thebibliography}{1}


\bibitem{malaney2004location} R. A. Malaney, ``A location enabled wireless security system,'' in \textit{Proc. IEEE GlobeCOM}, Nov. 2004, pp. 2196--2200.

\bibitem{vora2006secure} A. Vora, M. Nesterenko, ``Secure location verification using radio broadcast,'' \emph{IEEE Trans. on Dependable and Secure Computing}, vol. 3, no. 4, pp. 377--385, Oct. 2006.

\bibitem{chen2010detecting}  Y. Chen, J. Yang, W. Trappe, and R. P. Martin, ``Detecting and localizing identity-based attacks in wireless and sensor networks," \textit{IEEE Trans. Veh. Technol.}, vol. 59, no. 5, pp. 2418--2434, Jun. 2010.

\bibitem{chiang2012secure} J. T. Chiang, J. J. Haas, J. Choi, and Y. Hu ``Secure location verification using simultaneous multilateration,'' \emph{IEEE Trans. Wireless Commun.}, vol. 11, no. 2, pp. 584--591, Feb. 2012.

\bibitem{zekavat2011handbook} R. Zekavat and R. Buehrer, ``Handbook of Position Location: Theory, Practice and Advances,'' vol. 27. Wiley-IEEE Press, 2012.

\bibitem{yang2013detection} J. Yang, Y. Chen, and W. Trappe, and J. Cheng, ``Detection and localization of multiple spoofing attackers in wireless networks,¡± \emph{IEEE Trans. Parallel Distrib. Syst.}, vol. 24, no. 1, pp. 44--58, Jan. 2013.

\bibitem{yan2014signal} S. Yan, R. Malaney, I. Nevat, and G. Peters, ``Signal strength based location verification under spatially correlated shadowing,'' in \emph{Proc. IEEE ICC}, Jun. 2014, pp. 2617--2623.

\bibitem{yan2014optimal} S. Yan, R. Malaney, I. Nevat, and G. Peters, ``Optimal information-theoretic wireless location verification,'' \emph{IEEE Trans. Veh. Technol.}, vol. 63, no. 7, pp. 3410--3422, Sep. 2014.

\bibitem{malandrino2014verification} F. Malandrino, C. Borgiattino, C. Casetti, C.-F. Chiasserini, M.  Fiore, R. Sadao, ``Verification and inference of positions in vehicular networks through anonymous beaconing,'' \emph{IEEE Trans. Mobile Comput.}, vol. 13, no. 10, pp. 2415--2428, Oct. 2014.



\bibitem{leinmuller2005influence} T. Leinm$\ddot{u}$ller, E. Schoch, F. Kargl, and C. Maih$\ddot{o}$fer, ``Influence of falsified position data on geographic ad-hoc routing,'' in \emph{Proceedings of the second European Workshop on Security and Privacy in Ad hoc and Sensor Networks (ESAS)}, Jul. 2005, pp. 102--112.

\bibitem{capkun2010integrity} S. Capkun, M. Cagalj, G. Karame, and N.O. Tippenhauer, ``Integrity regions: authentication through presence in wireless networks'',  \emph{IEEE Trans. Mob. Comput.}, vol. 9, no. 11, pp. 1608--1621, Nov. 2010.

\bibitem{raya2007securing} M. Raya and J.-P. Hubaux, ``Securing vehicular ad hoc networks,'' \emph{J. Comput. Secur.}, vol. 15, no. 1, pp. 39--68, Jan. 2007.

\bibitem{yu2013detecting} B. Yu, C. Xu, and B. Xiao, ``Detecting sybil attacks in VANETs'', \emph{J. Parallel Distrib. Comput.}, vol. 73, no. 6, pp. 746--756, Jun. 2013.

 \bibitem{tang}  T. Zhang and L. Delgrossi, ``Vehicle Safety Communications: Protocols, Security, and Privacy,'' Wiley, 2012.




\bibitem{leinmuller2006position} T. Leinm$\ddot{u}$ller, E. Schoch, and F. Kargl, ``Position verification approaches for vehicular ad hoc networks,'' \textit{IEEE Wireless Commun.}, vol. 13, no. 5, pp. 16--21, Oct. 2006.

\bibitem{malaney2007wireless} R. Malaney, ``Wireless intrusion detection using tracking verification,'' in \emph{Proc. IEEE ICC}, Jun. 2007, pp.1558--1563.

\bibitem{yan2008providing} G. Yan, S. Olariu, and M. C. Weigle, ``Providing vanet security through active position detection,'' \emph{Computer Communications}, Vol. 31, no. 12, pp. 2883--2897, Jul. 2008.

\bibitem{yan2009providing} G. Yan, S. Olariu, and M. Weigle, ``Providing location security in vehicular ad hoc networks,'' \textit{IEEE Wireless Commun.}, vol. 16, no. 6, pp. 48--55, Dec. 2009.

\bibitem{hao2011cooperative} Y. Hao, J. Tang, and Y. Cheng, ``Cooperative sybil attack detection for position based applications in privacy preserved VANETs,'' in \emph{Proc. IEEE GlobeCOM}, Dec. 2011, pp. 1--5.

\bibitem{abumansoor2012a} O. Abumansoor, A. Boukerche, ``A secure cooperative approach for nonline-of-sight location verification in VANET," \emph{IEEE Trans. Veh. Technol.}, vol. 61, pp. 275--285, Jan. 2012.

\bibitem{zhang2012cooperative} P. Zhang, Z. Zhang, and A. Boukerche, ``Cooperative location verification for vehicular ad-hoc networks'', in \emph{Proc. IEEE ICC}, Jun. 2012, pp. 37--41.

\bibitem{fogue2014securing} M. Fogue, F. Martinez, P. Garrido, M. Fiore, C. Chiasserini, C. Casetti, J. Cano, C. Calafate, and P. Manzoni, ``Securing warning message dissemination in VANETs using cooperative neighbor position verification,'' \emph{IEEE Trans. Veh. Technol.}, DOI: 10.1109/TVT.2014.2344633.

\bibitem{yan2014location} S. Yan, R. Malaney, I. Nevat, and G. Peters, ``Location spoofing detection systems for VANETs by a single base station in Rician fading channels,'' submitted to \emph{IEEE VTC Spring}, arXiv:1410.2960, Oct. 2014.

\bibitem{meireles2010experi} R. Meireles, M. Boban, P. Steenkiste, O. Tonguz, and J. Barros, ``Experimental study on the impact of vehicular obstructions in VANETs,'' in \emph{Proc. IEEE Vehic. Net. Conf.}, Dec. 2010, pp. 338--345.

\bibitem{gozalves2012IEEE} J. Gozalves, M. Sepulcre, and R. Bauza, ``IEEE 802.11p vehicle to infrastructure communications in urban environments¡±, \emph{IEEE Commun. Mag.}, vol. 50, no. 5, pp. 176--183, May 2012.

\bibitem{taricco2011on} G. Taricco and E. Riegler, ``On the ergodic capacity of correlated Rician fading MIMO channels with interference,'' \emph{IEEE Trans. Inf. Theory}, vol. 57, no. 7, pp. 4123--4137, Jul. 2011.

\bibitem{prasad1993effects} R. Prasad and A. Kegel, ``Effects of Rician faded and lognormal shadowed signals on spectrum efficiency in microcellular radio,'' \emph{IEEE Trans. Veh. Technol.}, vol. 42, no. 3, pp. 274--281, Aug. 1993.

\bibitem{neyman1933problem} J. Neyman and E. Pearson, ``On the problem of the most efficient tests of statistical hypotheses,'' \textit{Phil. Trans. R. Soc. A}, vol. 231, pp. 289--337, Jan. 1933.

\bibitem{barkat2005signal} M. Barkat, \textit{Signal Detection and Estimation}. Boston, MA: Artech House, 2005.

\bibitem{eguchi2006interpreting} S. Eguchi and J. Copas, ``Interpreting Kullback-Leibler divergence with the Neyman-Pearson lemma,'' \emph{J. Multivar. Anal.}, vol. 97, no. 9, pp. 2034--2040, Oct. 2006.

\bibitem{kullback1951on} S. Kullback and R. A. Leibler, ``On information and sufficiency,'' \emph{Annals of Mathematical Statistics}, vol. 22, no. 1, pp. 79--86, Mar. 1951.


%
%




\bibitem{ioannides2005uniform} P. Ioannides and C. Balanis, ``Uniform circular arrays for smart antennas,'' \emph{IEEE Antennas Propag. Mag.}, vol. 47, pp. 192--206, Aug. 2005.


\end{thebibliography}
\end{document}